\documentclass[8pt]{article}
\usepackage[utf8]{inputenc}
\usepackage[a4paper,left=3cm,right=3cm,top=3.0cm,,bottom=3.0cm]{geometry}
\usepackage{amsmath}
\usepackage{amsfonts}
\usepackage{amssymb}
\usepackage[mathscr]{euscript}
\usepackage{xcolor}
\usepackage{graphicx}
\usepackage{pifont}
\usepackage{caption}
\usepackage{amsthm}
\usepackage{float}
\usepackage{multicol}
\usepackage{multirow}
\usepackage{pgf,tikz}
\usepackage{mathrsfs}
\usetikzlibrary{arrows}
\usepackage{pdflscape}
\usepackage{circuitikz}
\usetikzlibrary{calc}
\usepackage{adjustbox}
\usepackage{algorithm}
\usepackage{algpseudocode}
\usepackage{authblk}
\usepackage{hyperref}
\usepackage{wasysym}
\usepackage{enumitem}

\usepackage{caption,subcaption}
\usetikzlibrary{shapes.multipart}

\usepackage{pgfplots}
\pgfplotsset{compat=1.15}
\usepackage{mathrsfs}
\usetikzlibrary{arrows}

\DeclareMathOperator{\sgn}{sgn}

\DeclareMathOperator{\argmax}{argmax}

\title{On the Use of Cram\'{e}r-Rao Lower Bound for Least-Variance Circuit Parameters Identification of Li-ion Cells\\[17cm]}
\author{Vladimir Sovljanski\textsuperscript{a,*}, Mario Paolone\textsuperscript{a}\\
\href{mailto:vladimir.sovljanski@epfl.ch}{vladimir.sovljanski@epfl.ch}, \href{mailto:mario.paolone@epfl.ch}{mario.paolone@epfl.ch}}
\affil{\textsuperscript{a} Distributed Electrical Systems Laboratory, EPFL, 1015 Lausanne, Switzerland}

\date{}

\begin{document}

\begin{titlepage}
\maketitle
\noindent
\thispagestyle{empty}
\vspace{-45pt}
\begin{flushleft}
\large
\textsuperscript{*}Corresponding author\\
\end{flushleft}

\end{titlepage}

 \begin{abstract}
 
Electrochemical Impedance Spectroscopy (EIS) and Equivalent Circuit Models (ECMs) are widely used to characterize the impedance and estimate parameters of electrochemical systems such as batteries. 

We use a generic ECM with ten parameters grouped to model different frequency regions of the Li-ion cell’s impedance spectrum. We derive a noise covariance matrix from the measurement model and use it to assign weights for the fitting technique. The paper presents two formulations of the parameters identification problem. Using the properties of the ECM EIS spectra, we propose a method to initialize ECM parameters for the Complex Non-linear Least Squares (CNLS) technique.
 
The paper proposes a novel algorithm for designing the EIS experiments by applying the theory on Cram\'{e}r-Rao Lower Bound (CRLB) and Fisher Information Matrix (FIM) to the identification problem.
We show that contributions to the FIM elements strongly depend on the frequencies at which EIS is performed. Hence, the algorithm aims to adjust frequencies such that the most information about parameters is collected. This is done by minimizing the highest variance of ECM parameters defined by CRLB.
Results of a numerical experiment show that the estimator is efficient, and frequency adjustment leads to more accurate ECM parameters' identification.

 \end{abstract}

\pagebreak

\section{Introduction}
\setcounter{page}{1}

Estimating the Li-ion cell parameters is fundamental for designing the battery modules and packs, operating the battery and assessing its performance and degradation while serving its first or second life. The modeller may choose a specific parametric model depending on the application and desired level of detail and accuracy, which is opposed to computational complexity. Therefore, the selected model usually represents a trade-off between complexity, accuracy and computational efficiency. 
% In addition, depending on the available data, the modeller must ensure that the parameters are observable
In the literature, battery parametric models are classified as electrochemical and equivalent circuit models (ECMs) \cite{zhao_observability_2017}. Considering the constraints in computational power, efficiency, and data storage, the ECM remains the predominant choice among the available battery models employed in battery management systems (BMS). Together with Electrochemical Impedance Spectroscopy (EIS), ECMs can be used to characterize the parameters of the cells in the laboratory under controlled experimental conditions. In addition, the idea of implementing on-board EIS measurements within the BMS becomes more attractive thanks to its potential to estimate the battery impedance and its ECM parameters without dismantling the battery pack \cite{koseoglou_novel_2021} and to access the cells externally.

The cell and its parameters can be characterized in the time and frequency domains. In the time domain, the excitation signal is generic and can be given by the operation of the battery itself. This can lead to parameter identifiability issues if the excitation signal has only a few frequency components \cite{alavi_identifiability_2017}. In \cite{alavi_time-domain_2015}, authors used ECM with equivalent series resistance and one branch consisting of a resistor and constant-phase element (CPE) to model the cell and estimated parameters from time-domain data. 
In the frequency domain, the cell is usually excited with a small sinusoidal signal (e.g., current excitation) of a defined frequency while measuring the response (e.g., cell voltage) and remaining in a pseudo-linear regime around a specific operating point. This enables characterizing the cell impedance at different frequencies. An ECM is usually assigned to EIS spectra where the choice of ECM topology and elements depends on the EIS spectra shape and is decided by the modeller before the parameters identification. A critical review \cite{iurilli_use_2021} presents typical ECMs used in the literature for different types of cell chemistry. The authors also summarize how different parts of EIS spectra are modelled with ECM components.

The most common technique in the literature to estimate the ECM parameters from EIS data in the frequency domain is the Complex Non-linear Least Squares (CNLS) method, initially used in \cite{macdonald_analysis_1977} and applied to impedance data. The CNLS aims to minimize the weighted sum of squared mismatches between the measured and chosen (theoretical) model's impedance. The theory about CNLS, applied in this paper, can be found in a dedicated chapter of \cite{orazem_electrochemical_2017}. 
Due to the highly non-convex functions expressing the ECM's equivalent impedance, and consequently the non-convex objective function of CNLS, it is always necessary to initialize parameters before solving the optimization problem using different numerical techniques, for instance, Levenberg-Marquardt algorithm \cite{orazem_electrochemical_2017}, \cite{nocedal_numerical_2006}.

The literature often uses the Fisher Information Matrix (FIM) and the Cram\'{e}r-Rao lower bound (CRLB) for the experiment design. CRLB defines the minimum variance for an unbiased estimator.
In \cite{geuten_experimental_2007}, different optimal design strategies based on FIM are classified as A-, D- and E-optimal design \cite{goos_optimal_2011}, which refers to maximizing the trace, determinant and minimum eigenvalue of FIM, respectively.
Considering the optimal design of experiments on batteries, in \cite{pozzi_optimal_2018}, authors propose a method to maximize electrochemical single-particle model parameters identifiability by minimizing the trace of the inverse of FIM.
In \cite{schmidt_experiment-driven_2010}, authors use FIM to estimate the identifiability of electrochemical model parameters depending on measurements and improve the design of the experiments. 
Authors of \cite{pillai_optimizing_2022} proposed a method based on CRLB to optimize the current profiles used to estimate ECM parameters from online measurements.
In \cite{rothenberger_genetic_2015}, they maximize the determinant of FIM by adjusting the excitation current parameters and estimating the ECM parameters containing series resistance, one R$\parallel$C branch and an open-circuit voltage source in the time domain. 
Authors of \cite{du_information_2022} utilize FIM and CRLB to quantify the information carried by the measurements and ensure reliable ECM parameter identification. In this paper, we use the FIM to quantify the contributions that EIS measurements carry to the parameters of the ECM circuit that consists of the serial connection of pure resistor, constant phase element (CPE), two R$\parallel$CPE\footnote{Symbol $\parallel$ denotes parallel electrical connection.} branches (Zarc) and Warburg impedance, modelling the high-, mid- and low-frequency parts of the EIS spectrum.

Different weighting options are presented in \cite{macdonald_flexible_1987} as an alternative to the ideal case when the inverse of measured impedance variances is used. The alternative weighting strategies were usually applied when the measurement noise model was unknown or difficult to obtain. Nowadays, however, manufacturers of EIS measurement devices provide a detailed noise model. Still, in recent literature, authors use unity weights \cite{abaspour_robust_2022}, \cite{ospina_agudelo_identification_2019}. In \cite{troltzsch_characterizing_2006}, authors assign weights equal to the inverse of the squared impedance modules. Often, covariances between real and imaginary parts in CNLS formulated in Cartesian coordinates are neglected \cite{orazem_electrochemical_2017}. This paper weights mismatches using an inverse of an entirely derived covariance matrix corresponding to the measurement noise model.

 Initialization of the parameters for CNLS is often done by defining the physically meaningful ranges of parameters' values according to experience and choosing the values within the corresponding ranges. Authors of \cite{islam_unification_2020} initialize parameters for the Randles circuit by looking at some properties of the EIS spectra to extract the parameters' values quickly. However, for some parameters, the feasible range can be challenging to guess. In \cite{troltzsch_characterizing_2006}, a method is proposed for re-adjusting the feasible range for battery model parameters to automatize the fitting process. Still, the initial range for each parameter requires a manual adjustment. 
 In \cite{wu_battery_2023}, authors manually divide EIS spectra into regions corresponding to different parts of ECM and separately solve least-squares to obtain estimates. This approach might provide results for initialization rather than final estimates since, here, it is assumed that there is no overlapping between different elements in the spectra.

To the best of our knowledge, EIS measurements are usually performed at logarithmically distributed frequencies within the pre-defined frequency range and number of points per decade. When estimating the parameters, the variances of estimates are rarely compared to the theoretical minimum defined by CRLB. Therefore, this paper proposes the experimental design for characterizing the Li-ion battery impedance and accurately estimating ECM parameters from EIS measurements.

In view of the above state-of-the-art on ECM parameters identification from EIS data, the authors of this study identified two fundamental research questions: 
\begin{itemize}
    \item[i.] How to set frequencies that bring the most information in the parameters identification process?
    \item[ii.] How to best initialize the ECM parameters for solving the CNLS?
\end{itemize}

The key contributions of this work are summarized as follows:
\begin{itemize}
    \item[(1)] We first formulate the identification problem for Li-ion ECM parameters estimation from EIS measurements in a rigorous mathematical way in both polar and Cartesian coordinates. We derive a covariance matrix from the measurement model commonly provided by EIS instrument manufacturers. Its inverse weights the mismatch vectors between the measurements and model functions. We discuss the theory and analyze the connection between the obtained EIS spectra and elements of wide-band Li-ion ECM with ten parameters. This results in the development of a method for the automated initialization of the parameters that leverages the properties of the EIS spectra and parameters importance in different frequency regions.
    \item[(2)] We estimate the Li-ion cell ECM parameters using the weighted CNLS by solving the unconstrained minimization problem starting from the obtained initial parameter values. This provides the estimation algorithm to converge to the values with expected statistics, i.e., mean value and variance of the estimated parameters.
    \item[(3)] We derive the expressions for computing the FIM and CRLB for estimated parameters of Li-ion ECM from the EIS data in a general Gaussian form. This is used to quantify the best possible accuracy and to show that our estimator is efficient.
    \item[(4)] We improve the experimental design by proposing a novel algorithm to adjust the frequencies at which EIS should be performed to increase the accuracy of the estimates using the E-optimal design. We show that information about ECM parameters strongly depends on the frequencies at which EIS measurements are performed. The algorithm decides the frequency distribution over a predefined range by maximizing the FIM's lowest eigenvalue and minimizing the highest eigenvalue of its inverse matrix. EIS measurements at these frequencies contain more information about ECM parameters than traditionally used logarithmic span. Therefore, the final frequency adjustment leads to the EIS measurements from which ECM parameters can be estimated with lower variances and higher overall fitting accuracy.
\end{itemize}

This paper is organized as follows: in Section 2, we present the Li-ion battery ECM parameters estimation problem from EIS measurements in a rigorous way and discuss two different formulations and the problem of parameters' initialization, recall the theory of general Gaussian CRLB adapted to our problem and introduce a novel algorithm for improving the variance of the estimated parameters via CRLB. Section 3 presents the results of a numerical study where the parameters are estimated, and the overall estimation accuracy is improved after executing the proposed algorithm. Section 4 summarizes the main contributions and potentials of this work.

\section*{Keywords}
Li-ion batteries, electrochemical impedance spectroscopy, equivalent circuit models, parameters estimation, Cram\'{e}r-Rao lower bound, Fisher Information Matrix.

\pagebreak
\section{Method}

\subsection{General Aspects and Assumptions of Cells' Parameters Identification via EIS Measurements}
EIS is a powerful, non-invasive technique used to characterize Lithium-ion cells and generic electrochemical energy storage devices.  EIS is performed by exciting the battery cell with small sinusoidal signals (current in galvanostatic mode or voltage in potentiostatic mode) at different frequencies and measuring the cell's response (respectively, the voltage in galvanostatic mode or current in potentiostatic mode). Although electrochemical systems are nonlinear, for sufficiently small excitations (yet maintaining an acceptable signal-to-noise ratio), their response can be approximated as being pseudo-linear \cite{wang_electrochemical_2021}. Small excitation amplitudes ensure that, while performing the EIS, changes in the State-of-Charge (SoC) and cell temperature are negligible, i.e., $\Delta SoC \approx 0$ and $\Delta T\approx 0$.
Supposing that the voltage and current signals are $E(t) = E_0 + \Delta E \sin(\omega t)$ and $I(t) = I_0 + \Delta I \sin(\omega t - \phi)$, respectively, while operating around the point ($E_0, I_0$) on the nonlinear voltage-current curve, the complex impedance can be simply calculated as:
\begin{equation}
    \bar{Z}(\omega) = \dfrac{\Delta\bar{E}(\omega)}{\Delta\bar{I}(\omega)} = R(\omega) + jX(\omega).
\end{equation}
where $\Delta\bar{E}(\omega) = \Delta E(\omega) \angle 0$ and $\Delta\bar{I}(\omega) = \Delta I(\omega) \angle (-\phi)$ are corresponding phasors of sinusoidal perturbation signals $\Delta E \sin(\omega t)$ and $\Delta I \sin(\omega t - \phi)$.

The common method used to estimate the ECM parameters is the Complex Non-linear Least Square (CNLS). This is done by simultaneously fitting the real and imaginary parts of complex impedance measurements to fixed model functions which are usually non-linear. Various terms in the CNLS objective are weighted to quantify the heteroscedastic noise content of each observation.

The choice of elements comprising the ECM depends on the shape of the obtained complex impedance curve on the Nyquist plot and it is fixed by the modeller before the fitting procedure.
Once the ECM topology is fixed, the equivalent impedance $\hat{Z}(\boldsymbol{\theta},\omega)$ is expressed as a function of the parameter vector, $\boldsymbol{\theta}$, and angular frequency $\omega$.

In this paper, the assumptions are the following: (a) the small sine excitation signal is always chosen such that the system's response is pseudo-linear, (b) the measurement noise model is known, (c) we use a given parametric circuit model of a Li-ion battery cell.

The ECM parameters $\boldsymbol{\theta}$ are estimated by solving the following unconstrained optimization problem either in Cartesian or polar coordinates, written in the standard quadratic form:
\begin{equation}{\label{eq:objective_fun_prelim}}
    \boldsymbol{\hat{\theta}} = \underset{\boldsymbol{\theta}}{\text{argmin}}
    \left(\tilde{\mathcal{Z}}(\boldsymbol{\omega}) - {\mathcal{Z}}(\boldsymbol{\theta},\boldsymbol{\omega})\right)^\top {\boldsymbol{\tilde{Q}}}^{-1} \left(\tilde{\mathcal{Z}}(\boldsymbol{\omega}) - {\mathcal{Z}}(\boldsymbol{\theta},\boldsymbol{\omega})\right)
\end{equation}
where vector $\tilde{\mathcal{Z}}(\boldsymbol{\omega}) = \left[\tilde{z}_1,\dots,\tilde{z}_N\right]^\top$, with $\tilde{z}_i = \left[\tilde{R}_i, \tilde{X}_i\right]^\top$, $\tilde{R}_i=\Re(\tilde{Z}(\omega_i))$ and $\tilde{X}_i=\Im(\tilde{Z}(\omega_i))$ in Cartesian or $\tilde{z}_i = \left[\tilde{\rho}_i, \tilde{\varphi}_i\right]^\top$, $\tilde{\rho}_i=|\tilde{Z}(\omega_i)|$ and $\tilde{\varphi}_i=\arg(\tilde{Z}(\omega_i))$ in polar coordinates, contains separated real and imaginary parts in Cartesian, or modules and angles in polar coordinates, of impedances measured at $N$ frequencies contained in the vector $\boldsymbol{\omega} = \left[\omega_1,\dots,\omega_N\right]^\top$. On the other hand, $\mathcal{Z}(\boldsymbol{\theta},\boldsymbol{\omega})$ contains the impedance model (expressions for the real and imaginary part, or module and angle, of ECM impedance) as a function of ECM parameters $\boldsymbol{\theta}$ and evaluated at corresponding frequencies. The mismatches between the measurements and model are weighted using the inverse of the measurement covariance matrix, $\boldsymbol{\tilde{Q}}$.

Regardless of the formulation, the measurement covariance matrix $\boldsymbol{\tilde{Q}}$ has a block-diagonal form, $\boldsymbol{\tilde{Q}} = \text{diag}(\boldsymbol{\tilde{Q}}_1,\dots,\boldsymbol{\tilde{Q}}_N)$ since the measurements are independent. Therefore, the problem \eqref{eq:objective_fun_prelim} can be also written as:
\begin{equation}{\label{eq:objective_fun}}
    \boldsymbol{\hat{\theta}} = \underset{\boldsymbol{\theta}}{\text{argmin}}
    \sum_{i=1}^N
    \left(\tilde{z}_i - z_i(\boldsymbol{\theta})\right)^\top \boldsymbol{\tilde{Q}}_{i}^{-1} \left(\tilde{z}_i - z_i(\boldsymbol{\theta})\right)
\end{equation}

The objective function given by Eq. \eqref{eq:objective_fun} is generally highly non-convex due to the non-convexity of model functions ${{z}}_i(\boldsymbol{\theta})$. Hence, it can have multiple local minima, which motivates the need for good initialization of parameters that will be elaborated on later on in the paper.

\subsection{Measurement Model}\label{sec:measurementModel}

The EIS instrument's precision is often characterized by maximum relative error in magnitude, $\varepsilon_{\rho}$ and maximum absolute error in phase $\varepsilon_{\varphi}$. Manufacturers often provide the accuracy contour plots where one can read the values of $\varepsilon_{\rho}$ as a function of the impedance magnitude and frequency, $\varepsilon_{\rho}(f,|\bar{Z}|)$, while the maximum absolute error in phase is assumed to be constant, $\varepsilon_{\varphi} = const$. 

\begin{figure}[H]
    \centering
    \includegraphics[scale = 0.7]{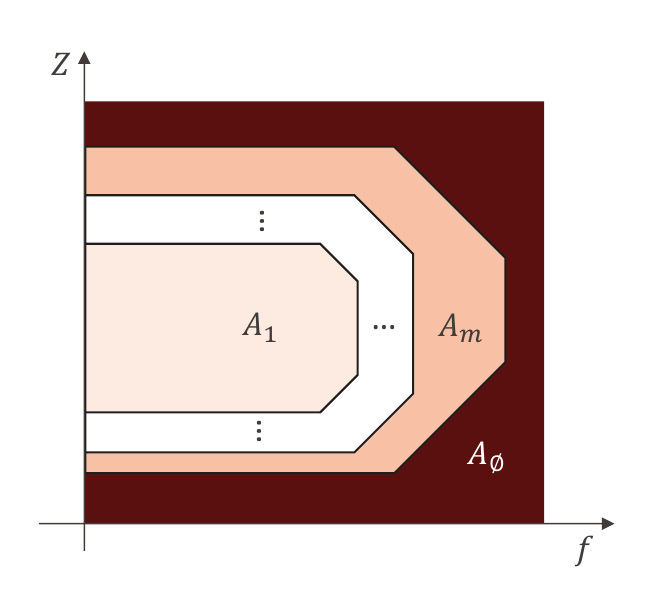}
    \caption{Conceptual accuracy contour plot of EIS instrument defining the areas $A_i, i = 1,\dots,m$ providing the values of the impedance relative magnitude error (undefined in the outer area $A_{\O}$ since $|\bar{Z}|$ exceeds the capability limits of an EIS instrument).}
    \label{fig:contour_plot}
\end{figure}
As shown in Fig. \ref{fig:contour_plot}, the accuracy contours divide the frequency-impedance plane into several areas. An accuracy contour plot area $A_i$ is defined as a set of points $(f,|\bar{Z}|)$  with $f > 0$ and $|\bar{Z}| > 0$, such that the EIS instrument measures the impedance magnitude $|\bar{Z}|$ with maximum magnitude relative error of $a_i \%$ at frequency $f$. Namely:
\begin{equation}
    \varepsilon_{\rho}(f,|\bar{Z}|) = 
    \begin{cases}
        a_1 \% &\text{if $(f,|\bar{Z}|)\in A_1$}\\
        % a_2 \% &\text{if $(f,|\bar{Z}|) \in A_2$}\\
        \vdots\\
        a_m \% &\text{if $(f,|\bar{Z}|) \in A_m$}\\
        \text{undefined}  &\text{if $(f,|\bar{Z}|) \in A_{\O}$}
    \end{cases}
\end{equation}
To express the standard deviations of the measured impedance module and phase, we assume that the measurement noise is Gaussian, unbiased, and lies within the specified maximum bound with a probability of 0.9973. Hence, we divide the maximum error value by 3. Therefore, standard deviations corresponding to magnitude and phase are:
\begin{equation}{\label{std_dev_rho_phi}}
         \sigma_{\rho}(\rho)  =  \dfrac{1}{3} \rho \cdot  \varepsilon_{\rho} \;\; \text{and} \;\; \sigma_{\varphi} =  \dfrac{1}{3} \cdot  \varepsilon_{\varphi} = const.
\end{equation}
Hence, for a true value of complex impedance $\bar{Z} = \rho \text{e}^{j\varphi}$, corresponding measured impedance is $\tilde{Z} = \tilde{\rho} \text{e}^{j\tilde{\varphi}} = (\rho + \Delta\rho) \text{e}^{j(\varphi + \Delta\varphi)} $ where the magnitude and phase errors are distributed as $\Delta\rho \sim \mathcal{N}(0,\frac{\rho  \varepsilon_{\rho}}{3} )$ and $\Delta\varphi \sim \mathcal{N}(0,\frac{\varepsilon_{\varphi}}{3})$, respectively.
\subsection{ECM Parameters Estimation via the CNLS in Polar Coordinates}
In polar coordinates, \eqref{eq:objective_fun} can be written as:
\begin{equation}{\label{eq:objective_fun_pol}}
    \boldsymbol{\hat{\theta}} = \underset{\boldsymbol{\theta}}{\text{argmin}}
    \sum_{i=1}^{N}
    \left(
    \begin{bmatrix}
    \tilde{\rho}_i\\
    \tilde{\varphi}_i
    \end{bmatrix}
    -
    \begin{bmatrix}
    {\rho}_i(\boldsymbol{\theta})\\
    {\varphi}_i(\boldsymbol{\theta})
    \end{bmatrix}
    \right)^\top
    \begin{bmatrix}
    \sigma_{\rho}^2(\rho_i)& 0\\ 
    0  & \sigma_{\varphi}^2
    \end{bmatrix}^{-1}
    \left(
    \begin{bmatrix}
    \tilde{\rho}_i\\
    \tilde{\varphi}_i
    \end{bmatrix}
    -
    \begin{bmatrix}
    {\rho}_i(\boldsymbol{\theta})\\
    {\varphi}_i(\boldsymbol{\theta})
    \end{bmatrix}
    \right).
\end{equation}
In this case, the covariance matrix has a pure diagonal form:
\begin{equation}
    \boldsymbol{\tilde{Q}} = \textbf{diag}(\sigma_{\rho_1}^2, \sigma_{\varphi}^2,\dots,\sigma_{\rho_N}^2, \sigma_{\varphi}^2)
\end{equation}
since it is assumed that stochastic errors in magnitude and phase are not correlated and, in addition, stochastic errors at different frequencies are not correlated with each other. The relationship between the measured impedance module and angle and real and imaginary parts is:
\begin{equation}
    |Z(\omega_i)| = \rho_i = \sqrt{R_i^2+ X_i^2}
\end{equation}

\begin{equation}
    \arg Z(\omega_i) = \varphi_i = \arctan\left({\dfrac{X_i}{R_i}}\right),\;\;\text{since}\;\; R_i > 0.
\end{equation}

Same relations hold for model functions in polar coordinates $\rho_i(\boldsymbol{\theta})$ and $\varphi_i(\boldsymbol{\theta})$ expressed in terms of model functions in Cartesian coordinates, $R_i(\boldsymbol{\theta})$ and $X_i(\boldsymbol{\theta})$.

\subsection{ECM Parameters Estimation via the CNLS in Cartesian Coordinates}
In Cartesian coordinates, the minimization problem \eqref{eq:objective_fun} becomes:
\begin{equation}{\label{eq:objective_fun_cart}}
    \boldsymbol{\hat{\theta}} = \underset{\boldsymbol{\theta}}{\text{argmin}}
    \sum_{i=1}^{N}
    \left(
    \begin{bmatrix}
    \tilde{R}_i\\
    \tilde{X}_i
    \end{bmatrix}
    -
    \begin{bmatrix}
    {R}_i(\boldsymbol{\theta})\\
    {X}_i(\boldsymbol{\theta})
    \end{bmatrix}
    \right)^\top
    \boldsymbol{\tilde{Q}}_i^{-1} 
    \left(
    \begin{bmatrix}
    \tilde{R}_i\\
    \tilde{X}_i
    \end{bmatrix}
    -
    \begin{bmatrix}
    {R}_i(\boldsymbol{\theta})\\
    {X}_i(\boldsymbol{\theta})
    \end{bmatrix}
    \right)
\end{equation}
Since the measurement errors are given in polar coordinates, while the measurements and model functions in Cartesian, we perform a transformation of coordinates (i.e., projection from polar to Cartesian coordinates), as in \cite{lerro_tracking_1993} and \cite{milano_static_2016}, to obtain $\boldsymbol{\tilde{Q}}_{i}$. Namely:

\begin{equation}
\boldsymbol{\tilde{Q}}_{i} = 
\begin{bmatrix}
\sigma^2(\tilde{R}_{i})    & \sigma(\tilde{R}_{i},\tilde{X}_{i})\\ 
\sigma(\tilde{R}_{i},\tilde{X}_{i})  & \sigma^2(\tilde{X}_{i})
\end{bmatrix}=
\begin{bmatrix}
\tilde{\alpha}_{i} &  \tilde{\beta}_{i}\\ 
\tilde{\beta}_{i}  &  \tilde{\gamma}_{i}
\end{bmatrix},
\end{equation}
whose elements are calculated using the measured impedance quantities (magnitude $\tilde{\rho}$ and phase $\tilde{\varphi}$):
\begin{align}
    \tilde{\alpha}_{i}   &=     \tilde{\rho}_i^2  e^{-2\sigma_\varphi^2}  
    \left[  \cos^2\tilde{\varphi}_i  (    \cosh(2\sigma_\varphi^2)  - \cosh(\sigma_\varphi^2)  ) + 
            \sin^2\tilde{\varphi}_i  (    \sinh(2\sigma_\varphi^2)  - \sinh(\sigma_\varphi^2) ) \right] \\
          &   +\sigma_{\rho_i}^2  e^{-2\sigma_\varphi^2}  
    \left[  \cos^2\tilde{\varphi}_i      (2\cosh(2\sigma_\varphi^2) - \cosh(\sigma_\varphi^2))     
          + \sin^2\tilde{\varphi}_i      (2\sinh(2\sigma_\varphi^2)   - \sinh(\sigma_\varphi^2)) \right]\nonumber\\
    \tilde{\beta}_{i}   &=     \tilde{\rho}_i^2  e^{-\sigma_\varphi^2}  
    \left[  \sin^2\tilde{\varphi}_i  (   \cosh(2\sigma_\varphi^2) - \cosh(\sigma_\varphi^2) ) 
          + \cos^2\tilde{\varphi}_i  (   \sinh(2\sigma_\varphi^2) - \sinh(\sigma_\varphi^2 ))    \right] \\
          &   +\sigma_{\rho_i}^2  e^{-\sigma_\varphi^2}  
    \left[  \sin^2\tilde{\varphi}_i     (2\cosh(2\sigma_\varphi^2) - \cosh(\sigma_\varphi^2))      
          + \cos^2\tilde{\varphi}_i     (2\sinh(2\sigma_\varphi^2) - \sinh(\sigma_\varphi^2))  \right]\nonumber\\
    \tilde{\gamma}_{i}  &= \sin\tilde{\varphi}_i\cos\tilde{\varphi}_i  e^{-4\sigma_\varphi^2} 
    \left[  \sigma_{\rho_i}^2 + 
            (\sigma_{\rho_i}^2 + \tilde{\rho}_i^2)  (1 - e^{\sigma_\varphi^2})  \right].
 \end{align}

\subsection{Accuracy Assessment of the Estimates}

Variances for the estimated parameters can be approximately calculated as diagonal elements of the matrix $\boldsymbol{\mathcal{A}}^{-1}$, calculated at values of estimated parameters, $\hat{\boldsymbol{\theta}}$ where, the element $(k,l)$ is expressed as:
\begin{equation}\label{eq:accuracy_alpha}
    \boldsymbol{\mathcal{A}}_{k,l}\approx
 \left\{
    \left[\dfrac{\partial{\mathcal{Z}}(\boldsymbol{\theta},\boldsymbol{\omega})}{\partial\theta_k}\right]^\top
    \boldsymbol{\tilde{Q}}^{-1}
    \left[\dfrac{\partial{\mathcal{Z}}(\boldsymbol{\theta},\boldsymbol{\omega})}{\partial\theta_l}\right]
    - 
    \left[{\tilde{\mathcal{Z}}(\boldsymbol{\omega}}) - {\mathcal{Z}}(\boldsymbol{\theta},\boldsymbol{\omega})\right]^\top
    \boldsymbol{\tilde{Q}}^{-1}
    \left[\dfrac{\partial^2{\mathcal{Z}}(\boldsymbol{\theta},\boldsymbol{\omega})}{\partial\theta_k\partial\theta_l}\right]
\right\}
\Bigg|_{\boldsymbol{\theta} = \hat{\boldsymbol{\theta}}}
\end{equation}
which corresponds to one-half of the corresponding $(k,l)$-th Hessian matrix element \cite{orazem_electrochemical_2017}.

\subsection{ECM Parameters Initial Guess}\label{sec:ECM_initial}
Since the objective function of the minimization problem given by Eq. \eqref{eq:objective_fun} is highly non-convex, the solution is very sensitive to the initial guess of the ECM parameters. Further, if the initial point is far from the true value, the solver is more likely to find a local minimum. Therefore, there is a need to improve the initial guess of parameters $\boldsymbol{\theta}$ and make the optimization solver more likely to find the estimates with the expectation equal to the actual true values. One possible approach can be to analyze the geometrical shapes of certain parts of EIS spectra. 

As a matter of fact, when connected, the spectra of each ECM element interfere and, therefore, their corresponding spectra overlap. Still, approximate values for ECM parameters can be extracted by fitting parts of the spectra to the expected geometrical shapes of each ECM branch. 
  
  \begin{minipage}[b]{0.55\textwidth}
    \centering
    \includegraphics[scale = 0.45]{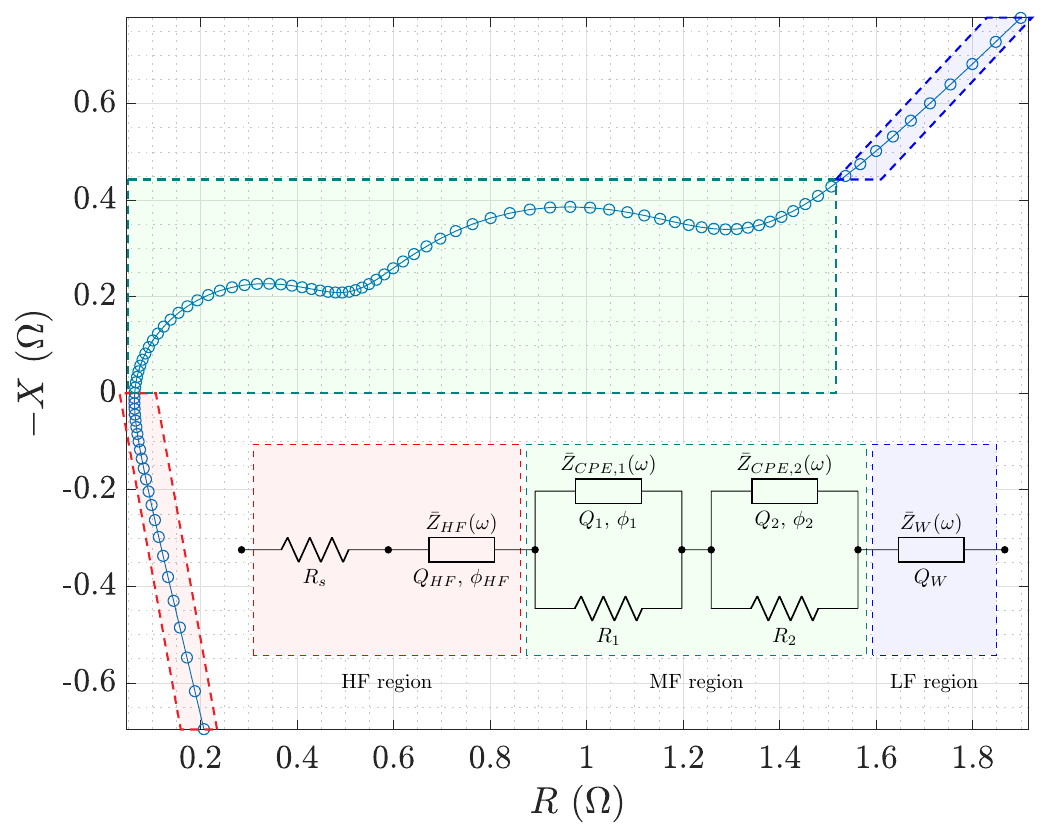}
    \captionof{figure}{Impedance spectrum of the Li-ion cell ECM. The parameters are indicated below each element.}
    \label{fig:SpectrumAndECM}
  \end{minipage}
  \hfill
  \begin{minipage}[b]{0.35\textwidth}
    \centering
    \small
        \begin{tabular}{|c|c|c|}
        \hline
        \textbf{Param.} & \textbf{Values} $\theta$ & \textbf{Unit} \\ \hline\hline
        $R_s$ & 3.80E-02 & $\Omega$ \\ \hline
        $Q_{HF}$ & 1.67E+04 & $\text{S}\cdot\text{s}^{-\varphi_{HF}}$ \\ \hline
        $\varphi_{HF}$ & -8.50E-01 & - \\ \hline
        $R_1$ & 4.50E-01 & $\Omega$ \\ \hline
        $Q_1$ & 2.00E-02 & $\text{S}\cdot\text{s}^{-\varphi_{1}}$ \\ \hline
        $\varphi_1$ & 9.00E-01 & - \\ \hline
        $R_2$ & 6.50E-01 & $\Omega$ \\ \hline
        $Q_2$ & 4.00E-01 & $\text{S}\cdot\text{s}^{-\varphi_{2}}$ \\ \hline
        $\varphi_2$ & 9.00E-01 & - \\ \hline
        $Q_W$ & 3.69E+00 & $\text{S}\cdot\text{s}^{-1/2}$ \\ \hline
        \end{tabular}
      \captionof{table}{ECM parameters' true values (inspired from \cite{ospina_agudelo_identification_2019}).}
      \label{tab:ECMparamValues}
    \end{minipage}

Fig. \ref{fig:SpectrumAndECM} shows the noiseless impedance spectrum and ECM used to model the Li-ion battery cell in this paper and highlights different frequency regions. Tab. \ref{tab:ECMparamValues} reports the ECM parameters' true values.
As known, the equivalent cell's impedance can be represented as a superposition of equivalent series resistance, CPE modelling the HF behaviour, a number (e.g., two) of Zarc elements (i.e., CPE in parallel to a resistor) and the Warburg impedance:
\begin{equation}
    \bar{Z}_{eq}(\boldsymbol{\theta};\omega) = R_s + \bar{Z}_{HF} + R_1\parallel\bar{Z}_{CPE,1} + R_2\parallel\bar{Z}_{CPE,2}+\bar{Z}_{W},
\end{equation}
where $z_1\parallel z_2 = \frac{z_1 z_2}{z_1 + z_2}$ is the equivalent impedance of parallel connection of $z_1$ and $z_2$. The expression for HF region CPE is
\begin{equation}
    \bar{Z}_{HF}(Q_{HF},\phi_{HF};\omega) = \frac{1}{(j\omega)^{\phi_{HF}}Q_{HF}},\;\;\phi_{HF}\in\left[-1,0\right),
\end{equation}
while for two CPEs within the MF region, we have:
\begin{equation}
    \bar{Z}_{CPE,i}(Q_{i},\phi_{i};\omega) = \frac{1}{(j\omega)^{\phi_{i}}Q_{i}},\;\;\phi_{i}\in\left(0,1\right], \;\; i\in\{1,2\}.
\end{equation}
The Warburg impedance is a special case of CPE with the exponent $\phi$ equal to $\frac{1}{2}$:
\begin{equation}
    \bar{Z}_{W}(Q_{W};\omega) = \frac{1}{\sqrt{j\omega}\cdot Q_{W}}.
\end{equation}
Separating the real $\Re\{\bar{Z}_{eq}(\boldsymbol{\theta};\omega)\}={R}_{eq}(\boldsymbol{\theta},\omega)$ and imaginary $\Im\{\bar{Z}_{eq}(\boldsymbol{\theta};\omega)\}={X}_{eq}(\boldsymbol{\theta},\omega)$ parts of the equivalent impedance yields:

${R_i}(\boldsymbol{\theta})=\Re\{{Z}_i(\boldsymbol{\theta})\}$ and ${X_i}(\boldsymbol{\theta})=\Im\{{Z}_i(\boldsymbol{\theta})\}$

\begin{align}
    {R}_{eq}(\boldsymbol{\theta},\omega) \label{eq:zeq_real}
    &= R_s
    &+& \tfrac{1}{\omega^{\phi_{HF}}Q_{HF}}\cos\delta_{HF}
    + \sum_{i=1}^2 \tfrac{R_i + R_i^2 Q_i \omega^{\phi_i} \cos\delta_{i}}{1 + 2 R_i Q_i \omega^{\phi_i} \cos\delta_{i} + \left(R_i Q_i \omega^{\phi_i}\right)^2}
    + \tfrac{1}{Q_W\sqrt{2\omega}}\\
    {X}_{eq}(\boldsymbol{\theta},\omega) \label{eq:zeq_imag}
    &=
    &-&\tfrac{1}{\omega^{\phi_{HF}}Q_{HF}}\sin\delta_{HF}
    - \sum_{i=1}^2 \tfrac{R_i^2 Q_i \omega^{\phi_i} \cos\delta_{i}}{1 + 2 R_i Q_i \omega^{\phi_i} \cos\delta_{i} + \left(R_i Q_i \omega^{\phi_i}\right)^2}
    - \tfrac{1}{Q_W\sqrt{2\omega}}
\end{align}
where $\delta_{HF}=\frac{\phi_{HF} \pi}{2}$ and $\delta_{i}=\frac{\phi_i \pi}{2}$, $i\in\{1,2\}$.
\subsubsection{On the Geometry of ECM Elements Spectra}
Individually, all the ECM electrical elements, with fixed parameters, produce spectra with a known locus $\mathcal{L}$ governed by the expressions for its impedance:
\begin{equation}
    \mathcal{L} = \{(R(\omega),-X(\omega)): R(\omega) = \Re(\bar{Z}(\omega)) \wedge X(\omega)=\Im(\bar{Z}(\omega))\wedge \omega\in \left(0,\infty\right)\}.
\end{equation}

Note that we always consider the Nyquist plot with axes indicating $R$ and $-X$ as a usual convention for plotting the EIS measurements.

The locus of the resistance $R_s$ is simply a singleton $(R_s, 0)$ at the Nyquist plot. Adding a pure resistance in series to the ECM translates the entire spectrum in the direction of the horizontal axis by the value of the resistance.

The locus of CPE with parameters $Q$ and $\phi$ represents a line starting from the origin and an angular coefficient of $\frac{\pi\phi}{2}$.  

By adding the resistance $R$ in parallel to the CPE (i.e., the Zarc element), the locus represents a part of a depressed circle crossing the real axis at $(0,0)$ and $(R,0)$, centred at $\left(\frac{R}{2},-\frac{R}{2}\cot\frac{\pi\phi}{2}\right)$, with a radius $ r = \frac{R}{2\sin\frac{\pi\phi}{2}}$ and $\Im{\{\bar{Z}_{arc}\}} < 0$.

The CPE can be seen as a special case of a Zarc element for which $R\rightarrow\infty$ and, therefore, the locus of the CPE impedance is tangent to Zarc if their exponents are equal.

Different elements dominate in different frequency regions as indicated in Fig. \ref{fig:SpectrumAndECM}. 
However, the spectra produced by individual elements interfere. This results in the equivalent superposed spectrum for which the locus of different frequency regions deviates from the ideal geometrical shapes (as was the case for individual ECM elements). Still, the approximate values for ECM parameters can be extracted and used to define each ECM parameter's initial guess and reasonable interval of feasibility. In such a way, we reduce the solution space of the estimation problem.

Theoretically, as $\omega\rightarrow 0$, the Warburg impedance modelling the LF region dominates the EIS spectrum while the influence of HF and MF parts is negligible. Since:
\begin{equation}
    k_{LF} = \lim_{\omega\rightarrow 0} \dfrac{\Im\{\bar{Z}_{eq}(\boldsymbol{\theta};\omega)\}}{\Re\{\bar{Z}_{eq}(\boldsymbol{\theta};\omega)\}}= -1
\end{equation}
and
\begin{equation}
    n_{LF} = \lim_{\omega\rightarrow 0}\left[\Im\{\bar{Z}_{eq}(\boldsymbol{\theta};\omega)\} - k_{LF}\Re\{\bar{Z}_{eq}(\boldsymbol{\theta};\omega)\}\right] = R_s + R_1 + R_2 = R_{\Sigma},
\end{equation}
on the Nyquist plot $(R,-X)$, the spectrum asymptotically approaches the line $ -X = R + R_{\Sigma}$.

Similarly, as $\omega\rightarrow\infty$, the HF spectrum asymptotically approach the line $X = k_{HF} R + n_{HF}$ where:
\begin{equation}
    k_{HF} = \lim_{\omega\rightarrow \infty} \dfrac{\Im\{\bar{Z}_{eq}(\boldsymbol{\theta};\omega)\}}{\Re\{\bar{Z}_{eq}(\boldsymbol{\theta};\omega)\}} = 
    \lim_{\omega\rightarrow \infty} \dfrac{\Im\{\bar{Z}_{eq}(\boldsymbol{\theta};\omega)\}}{\Re\{\bar{Z}_{eq}(\boldsymbol{\theta};\omega)\} - R_s} = -\tan\left(\frac{\pi\phi_{HF}}{2}\right)
\end{equation}
and
\begin{equation}\label{eq:nHFandRs}
    n_{HF} = \lim_{\omega\rightarrow \infty}\left[\Im\{\bar{Z}_{eq}(\boldsymbol{\theta};\omega)\} - k_{HF}\Re\{\bar{Z}_{eq}(\boldsymbol{\theta};\omega)\}\right] =R_s\tan\left(\frac{\pi\phi_{HF}}{2}\right).
\end{equation}

\begin{figure}[H]
    \centering
    \includegraphics[scale = 0.6]{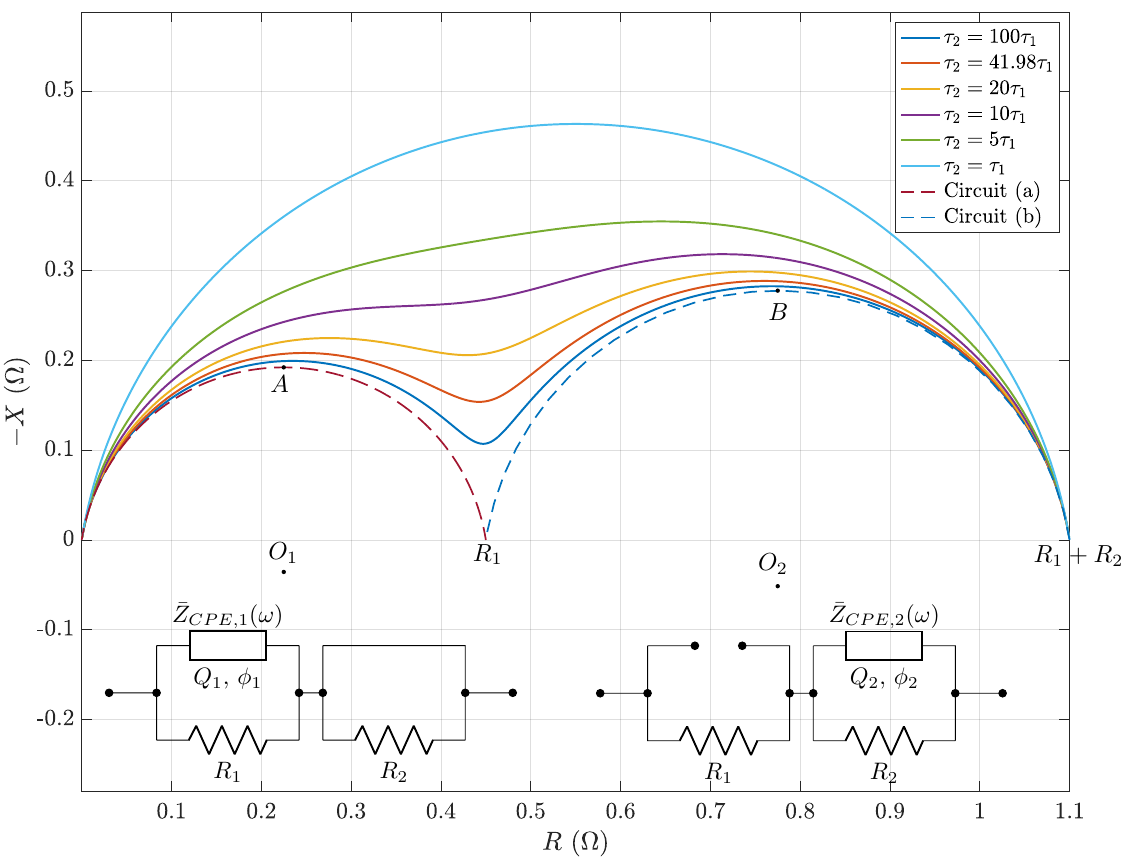}
    \caption{Equivalent MF region spectrum cleaned from the interference produced by $\bar{Z}_{HF}$ and $\bar{Z}_{LF}$ for different ratios of the time constants of $\bar{Z}_{arc,1}$ and $\bar{Z}_{arc,2}$ (cell's parameters $R_1$, $Q_1$, $\phi_1$, $R_2$ and $\phi_2$ true values are those reported in Tab.~\ref{tab:ECMparamValues}, while parameter $Q_2$ changes its values to achieve a desired $\frac{\tau_2}{\tau_1}$ ratio).}
    \label{fig:TwoCPEs}
\end{figure}

We assume that two Zarc elements have different time constants $\tau_1$ and $\tau_2$, $\tau_1 < \tau_2$, defined by
    \begin{equation}
        \tau_1 = (R_1 Q_1)^\frac{1}{\phi_1}\;\;
        \text{and}\;\;
        \tau_2 = (R_2 Q_2)^\frac{1}{\phi_2}.
    \end{equation}
An example of the spectra of MF region produced only by two Zarc elements is shown in Fig. \ref{fig:TwoCPEs} for different time constants of Zarcs by varying only the parameter $Q_2$. For parameters' values reported in Tab. \ref{tab:ECMparamValues}, $\tau_2 = 41.98 \tau_1$.
The superposed MF spectrum of the two Zarcs is positioned above parts of two depressed semi-circles located above the horizontal axis, centred at $O_1$ and $O_2$ with radii $\frac{R_1}{2\sin\frac{\pi\phi_1}{2}}$ and $\frac{R_2}{2\sin\frac{\pi\phi_2}{2}}$, respectively. Both represent the locus of the MF region ECM with two Zarcs connected in series: in the first one CPE2 is short-circuited, while in the second one, CPE1 is replaced by an open-circuit. 
The interference of two Zarc elements spectra depends on the ratio between time constants, $\frac{\tau_2}{\tau_1}$. The greater the ratio, the closer is the superposed spectra to these depressed semi-circles. For these two Zarcs, the characteristic angular frequencies are given by
\begin{equation}\label{eq:CharactFreqs}
        \omega_{c1} = \dfrac{1}{(R_1 Q_1)^\frac{1}{\phi_1}}\;\;
        \text{and}\;\;
        \omega_{c2} = \dfrac{1}{(R_2 Q_2)^\frac{1}{\phi_2}}.
\end{equation}
The MF impedance (excluding the HF and LF impedances) is
\begin{equation}
    \bar{Z}_{MF}(\boldsymbol{\theta};\omega) = R_1\parallel\bar{Z}_{CPE,1} +  R_2\parallel\bar{Z}_{CPE,2} = \dfrac{R_1}{1 + (j\omega)^{\phi_1}R_1 Q_1} + \dfrac{R_2}{1 + (j\omega)^{\phi_2}R_2 Q_2}
\end{equation}
and at $\omega_{c1}$ and $\omega_{c2}$ it has the following values:
\begin{equation}
    \bar{Z}_{MF}(\boldsymbol{\theta}_{MF};\omega_{c1}) = \dfrac{R_1}{1 + j^{\phi_1}} + \dfrac{R_2}{1+j^{\phi_2}\left(\frac{\tau_2}{\tau_1}\right)^{\phi_2}}
\end{equation}
\begin{equation}
    \bar{Z}_{MF}(\boldsymbol{\theta}_{MF};\omega_{c2}) =  \dfrac{R_1}{1+j^{\phi_1}\left(\frac{\tau_1}{\tau_2}\right)^{\phi_1}} + \dfrac{R_2}{1 + j^{\phi_2}}.
\end{equation}
If $\tau_1 << \tau_2$ (i.e., the processes inside the cell modelled with Zarc elements can be well distinguished) we have:
\begin{equation}\label{eq:MFeq1}
    \bar{Z}_{MF}(\boldsymbol{\theta}_{MF};\omega_{c1}) \approx \dfrac{R_1}{1 + j^{\phi_1}} + R_2(0 + j0) = \dfrac{R_1}{2} - j\dfrac{R_1}{2}\tan\frac{\pi\phi_1}{4},
\end{equation}
\begin{equation}\label{eq:MFeq2}
    \bar{Z}_{MF}(\boldsymbol{\theta}_{MF};\omega_{c2}) \approx  R_1(1 + j0) + \dfrac{R_2}{1 + j^{\phi_2}} = R_1 + \dfrac{R_2}{2} - j\dfrac{R_2}{2}\tan\frac{\pi\phi_2}{4}
\end{equation}
which are exactly the peaks of the two depressed semi-circles defined by Zarc1 and Zarc2 elements while having the CPE2 and CPE1 replaced by a short-circuit and open-circuit, respectively. In Fig.~\ref{fig:TwoCPEs}, these two peaks are denoted as $A$ and $B$.

\subsubsection{Parameters Initialization Procedure}\label{sec:InitializationLF_HF_MF}
In what follows, we propose a detailed procedure to initialize the ECM parameters for the CNLS using the discussed properties of the spectra. First, the LF and HF parameters are estimated from two ends of the EIS spectrum.

\textbf{Fitting the LF and HF spectra to a line.} 
We first choose the number of points $N_{LF}$ and $N_{HF}$ used for fitting such that the residuals between the selected points of the LF and HF spectra and the fitted line are normally distributed while respecting the maximum relative error in magnitude and maximum absolute error in phase as defined in Section \ref{sec:measurementModel}. This ensures that the points belong to the part of the spectrum where the Warburg impedance and CPE used to model the LF and HF spectrum dominate, respectively. The procedure is described in the following pseudo-code.

\algdef{SE}[DOWHILE]{Do}{doWhile}{\algorithmicdo}[1]{\algorithmicwhile\ #1} % do-while
\begin{algorithm}[H]
\caption{Determining $N_{xF}$, $k_{xF}$ and $n_{xF}$, $x\in\{L,H\}$}\label{alg:HF_LF_coef}
\begin{algorithmic}[1]
\State \textbf{Initialize} $N_{xF} \gets 1$, $\varepsilon_{\rho} = \varepsilon_{\rho}(f,|Z|)$, $\varepsilon_\varphi = const$;
\State Sort the EIS data on $f$ in decreasing (for HF) or increasing order (for LF);
  \Do
    \State $N_{xF} \gets N_{xF} + 1$;
    \State Obtain $k_{xF}$ and $n_{xF}$ by solving:
    \begin{equation}\label{eq:HFLFspectraArgmin}
        \begin{aligned}
        \underset{k_{xF},n_{xF}}{\text{argmin}} \quad & \sum_{i=1}^{N_{xF}}d_i^2\\
        \textrm{s.t.} \quad & d_i = \left|\tilde{Z}_i-{Z}_i^\mathcal{L}\right|\\
                    & \mathcal{L}(R_i) = k_{xF}{R}_i + n_{xF}    \\
                    & k_{xF} \in \begin{cases}
                        \{-1\},                 & x = L\\
                        \left[0,+\infty\right), & x = H
                    \end{cases}\\
                    & {Z}_i^\mathcal{L} = \textbf{proj}_\mathcal{L} \tilde{Z}_i
        \end{aligned}
    \end{equation}
    \State $ \tilde{\varepsilon}_{\rho_i} = \left| \frac{|Z_i^\mathcal{L}| - |\tilde{Z_i}|}{|Z_i^\mathcal{L}|} \right| $, \quad $ \tilde{\varepsilon}_{\varphi_i} = \left| \arg{Z_i^\mathcal{L}} - \arg{\tilde{Z_i}} \right| $
  \doWhile{$ \tilde{\varepsilon}_{\rho_i} \leqslant \varepsilon_\rho(f_i,{\tilde{\rho}_i}) $ \textbf{and} $\tilde{\varepsilon}_{\varphi_i} \leqslant \varepsilon_\varphi, \forall i \in\{1,\dots, N_{xF}\}$ }% <--- use \doWhile for the "while" at the end
\end{algorithmic}
\end{algorithm}

The algorithm initializes the starting number of data points $N_{xF}$ and the measurement model through $\varepsilon_\rho$ and $\varepsilon_\varphi$. 
At each iteration, it increases the number of data points used for fitting and finds the slope and the interception by fitting $N_{xF}$ consecutive end-points of the spectra (either from the LF or HF) to a line. Namely, it solves a simple least-squares problem \eqref{eq:HFLFspectraArgmin} by minimizing the sum of squared distances $d_i$ between the measurements $\tilde{Z_i}$ and their corresponding normal projections (in Algorithm ~\ref{alg:HF_LF_coef}, $\textbf{proj}_\mathcal{L}$ denotes the projection operator) to a line $\mathcal{L}$, denoted as $Z_i^\mathcal{L}$ for $i=1,\dots, N$. It uses the property that the HF and LF spectra asymptotically approach the line with slope $k_{HF}\in\left[0,+\infty\right)$ and $k_{LF} = -1$, respectively. The algorithm then computes all the relative errors in magnitude and absolute errors in phase between the measured points and corresponding fitted points on the line. It terminates if any of the errors exceed the defined error limits.

\newpage
\textbf{Initialization of LF parameter $Q_W$.} 
Starting from the lowest frequency, $N_{LF}$ points are used to fit the spectra to a line $X(R) = k_{LF}R+n_{LF}$, specifically $-X = R+R_\Sigma$ for which:
\begin{equation}
    \tilde{R}_i + j\tilde{X}_i \approx R_\Sigma + \bar{Z}_W = R_\Sigma + \dfrac{1}{Q_W\sqrt{\omega_i}}e^{-j\frac{\pi}{4}},\;\; i = 1,\dots,N_{LF}.
\end{equation}
Therefore, $Q_W$ can be initialized with the mean value using the expressions for $Q_W$ from both real and imaginary parts of these selected points: 
\begin{equation}
    Q_W^0 = \dfrac{1}{2N_{LF}}\left(\sum_{i=1}^{N_{LF}} \dfrac{1}{\sqrt{2\omega_i}(\tilde{R}_i - R_\Sigma)} - \sum_{i=1}^{N_{LF}}\dfrac{1}{\sqrt{2\omega_i} \tilde{X}_i}\right),
\end{equation}
where $R_\Sigma$ is the interception of the fitted line with the horizontal axis. Therefore, fitting the LF part of the spectra has provided us with the initial guess for $Q_W$ and the sum of all the resistances of the ECM, $R_\Sigma = R_s + R_1 + R_2$.

\textbf{Initialization of HF parameters $R_s$, $Q_{HF}$ and $\phi_{HF}$.}
Choosing the $N_{HF}$ points with the highest frequencies, for which again the residuals are normally distributed, we fit the HF spectra to a line $-X = k_{HF}R+n_{HF}$ where the chosen points are expressed as:
\begin{equation}\label{eq:NHFImagPart}
    \tilde{R}_i + j\tilde{X}_i \approx R_s + \bar{Z}_{HF} = R_s + \dfrac{1}{\omega_i^{\phi_{HF}}Q_{HF}}e^{-j\frac{\pi\phi_{HF}}{2}},\;\; i = 1,\dots,N_{HF}.
\end{equation}
On obtaining the slope $k_{HF}$, the exponent $\phi_{HF}$ is
\begin{equation}
    \phi_{HF}^0 \approx \dfrac{\arctan k_{HF}}{\pi/2}.
\end{equation}
Now, $R_s$ can be obtained by simply using \eqref{eq:nHFandRs} as
\begin{equation}
    R_s^0 \approx \dfrac{n_{HF}}{\tan\left(\dfrac{\pi\phi_{HF}^0}{2}\right)},
\end{equation}
and $Q_{HF}$ can be initialized by averaging the expressions for $Q_{HF}$ from the real and imaginary part of \eqref{eq:NHFImagPart} evaluated at $\tilde{R}_i$ and $\tilde{X}_i$ over the chosen $N_{HF}$ points:
\begin{equation}
    Q_{HF}^0 = \dfrac{1}{2 N_{HF}}\left( \sum_{i = 1}^{N_{HF}} \dfrac{1}{\left(\tilde{R}_i - R_s^0\right) \omega_i^{\phi_{HF}^0}}\cos\frac{\pi\phi_{HF}^0}{2} - \sum_{i = 1}^{N_{HF}} \dfrac{1}{\tilde{X}_i\omega_i^{\phi_{HF}^0}}\sin\frac{\pi\phi_{HF}^0}{2}\right).
\end{equation}

\textbf{Initialization of MF parameters $R_1$, $Q_1$, $\phi_1$ and $R_2$, $Q_2$, $\phi_2$}.
To initialize the MF parameters, we subtract the LF and HF spectra from the original one using their initial values:
\begin{equation}
    \bar{Z}_{MF}(\boldsymbol{\theta}_{MF}^0; \omega) \approx \bar{Z}_{eq}(\boldsymbol{\theta}; \omega) - \bar{Z}_{LF}(\boldsymbol{\theta}_{LF}^0; \omega) - \bar{Z}_{HF}(\boldsymbol{\theta}_{HF}^0; \omega).
\end{equation}
We can then find two peak points of the MF spectra corresponding to two Zarcs at their characteristic frequencies $\omega_{c1}$ and $\omega_{c2}$, we use Eq.~\eqref{eq:MFeq1} and \eqref{eq:MFeq2} to initialize the parameters:
\begin{equation}
    R_1^0 = 2\cdot\Re\{\bar{Z}_{MF}(\boldsymbol{\theta};\omega_{c1})\}\;\;\text{and}\;\;
    \phi_1^0 = \dfrac{4}{\pi}\arctan\left({-\dfrac{2\Im\{\bar{Z}_{MF}(\boldsymbol{\theta};\omega_{c1})\}}{R_1^0}}\right)
\end{equation}
and similarly
\begin{equation}
    R_2^0 = 2\cdot(\Re\{\bar{Z}_{MF}(\boldsymbol{\theta};\omega_{c2})\} - R_1^0)\;\;\text{and}\;\;
    \phi_2^0 = \dfrac{4}{\pi}\arctan\left({-\dfrac{2\Im\{\bar{Z}_{MF}(\boldsymbol{\theta};\omega_{c2})\}}{R_2^0}}\right).
\end{equation}
The remaining parameters $Q_1$ and $Q_2$ can be initialized by expressing them from \eqref{eq:CharactFreqs}:
    \begin{equation}
        Q_{1}^0 = \dfrac{1}{(\omega_{c1})^{\phi_1^0} R_1^0}\;\;
        \text{and}\;\;
        Q_{2}^0 = \dfrac{1}{(\omega_{c2})^{\phi_2^0} R_2^0}.
    \end{equation}

\pagebreak

\subsection{General Gaussian CRLB for Estimated ECM Parameters}
This section assesses the CRLB to find and compute the best possible CNLS estimator's variance. Since, in polar and Cartesian coordinates, the covariance matrix can be expressed as a function of parameters, $\boldsymbol{Q}(\boldsymbol{\theta}) = \text{diag}(\boldsymbol{Q}_1(\boldsymbol{\theta}),\dots,\boldsymbol{Q}_N(\boldsymbol{\theta}))$, a general Gaussian CRLB can be derived.

First, we assume that the probability density function of measurement mismatches is normal. I.e., the probability density function is given by the following expression:
    \begin{equation}
        f(\tilde{\mathcal{Z}} | \boldsymbol{\theta}, \boldsymbol{\omega}) = \dfrac{1}{\sqrt{(2\pi)^N \det \boldsymbol{{Q}}(\boldsymbol{\theta}) }} \exp\left[-\dfrac{1}{2} \boldsymbol{\varepsilon}^\top(\boldsymbol{\theta},\boldsymbol{\omega})\,\boldsymbol{{Q}}^{-1}(\boldsymbol{\theta})\,\boldsymbol{\varepsilon}(\boldsymbol{\theta},\boldsymbol{\omega})\right],
    \end{equation}
    where 
    $\boldsymbol{\varepsilon}(\boldsymbol{\theta},\boldsymbol{\omega}) = \tilde{\mathcal{Z}}(\boldsymbol{\omega}) - {\mathcal{Z}}(\boldsymbol{\theta},\boldsymbol{\omega})$
    is the vector containing mismatches between the measurements and model function at frequencies $\boldsymbol{\omega}$.
    This enables us to find the Log-likelihood function $L(\boldsymbol{\theta} | \tilde{\mathcal{Z}}) = \ln f(\tilde{\mathcal{Z}}|\boldsymbol{\theta})$, expressed as follows:
    \begin{equation}
        L(\boldsymbol{\theta} | \tilde{\mathcal{Z}}) = \ln f(\tilde{\mathcal{Z}}|\boldsymbol{\theta})
    \end{equation}
    \begin{equation}
        L(\boldsymbol{\theta},\boldsymbol{\omega} | \tilde{\mathcal{Z}}) = -\dfrac{N}{2}\ln 2\pi - \dfrac{1}{2}\ln\det \boldsymbol{Q}(\boldsymbol{\theta}) - \dfrac{1}{2}\boldsymbol{\varepsilon}^\top(\boldsymbol{\theta},\boldsymbol{\omega})\,\boldsymbol{Q}^{-1}(\boldsymbol{\theta}) \,\boldsymbol{\varepsilon}(\boldsymbol{\theta},\boldsymbol{\omega})
    \end{equation}
    and to compute its first-order partial derivatives with respect to every parameter $\theta_j$, $j = 1,\dots, M$:
    
    \begin{align*}
        \frac{\partial L(\boldsymbol{\theta}, \boldsymbol{\omega} | \tilde{\mathcal{Z}})}{\partial \theta_j} & = 
        -\frac{1}{2}\dfrac{\partial}{\partial\theta_j} \ln \det \boldsymbol{Q}(\boldsymbol{\theta})
        -\frac{1}{2}\dfrac{\partial}{\partial\theta_j}\left(\boldsymbol{\varepsilon}^\top(\boldsymbol{\theta},\boldsymbol{\omega})\,\boldsymbol{Q}^{-1}(\boldsymbol{\theta})\right)\\
        & = 
        -\frac{1}{2}\text{tr}\left(\boldsymbol{Q}^{-1}(\boldsymbol{\theta}) \frac{\partial \boldsymbol{Q}(\boldsymbol{\theta})}{\partial\theta_k} \right)
        +\frac{\partial \mathcal{Z}^\top(\boldsymbol{\theta},\boldsymbol{\omega})}{\partial\theta_k}\,\boldsymbol{Q}^{-1}(\boldsymbol{\theta}) \boldsymbol{\varepsilon}(\boldsymbol{\theta},\boldsymbol{\omega})
        -\frac{1}{2} \boldsymbol{\varepsilon}^\top(\boldsymbol{\theta},\boldsymbol{\omega}) \frac{\partial \boldsymbol{Q}(\boldsymbol{\theta})}{\partial\theta_k} \boldsymbol{\varepsilon}(\boldsymbol{\theta},\boldsymbol{\omega}).
    \end{align*}

    The $(k,l)$-th element of the FIM, $ \boldsymbol{\mathcal{F}} \in \mathbb{R}^{M \times M}$, as a function of parameter vector $\boldsymbol{\theta}$ and angular frequency $\omega$, is equal to the expected value of the product between partial derivatives of the Log-likelihood function with respect to $k$-th and $l$-th parameter:
    \begin{equation}
        \boldsymbol{\mathcal{F}}_{k,l}(\boldsymbol{\theta}, \boldsymbol{\omega}) = \mathbb{E}\left[ \dfrac{\partial L(\boldsymbol{\theta}, \boldsymbol{\omega} | \tilde{\mathcal{Z}})}{\partial \theta_k} \dfrac{\partial L(\boldsymbol{\theta}, \boldsymbol{\omega} | \tilde{\mathcal{Z}})}{\partial \theta_l}\right].
    \end{equation}

The expression for $(k,l)$-th element of the FIM can be written in compact form as follows \cite{kay_fundamentals_1993}:
\begin{equation}\label{eq:CRLB_exp}
[ \boldsymbol{\mathcal{F}}(\boldsymbol{\theta},\boldsymbol{\omega})]_{k,l} =  \dfrac{\partial{\mathcal{Z}}^\top(\boldsymbol{\theta},\boldsymbol{\omega})}{\partial\theta_k} \boldsymbol{Q}^{-1}(\boldsymbol{\theta})\dfrac{\partial{\mathcal{Z}}(\boldsymbol{\theta},\boldsymbol{\omega})}{\partial\theta_l}
+\dfrac{1}{2}\text{tr}
\left(
\boldsymbol{Q}^{-1}(\boldsymbol{\theta})
\cdot\dfrac{\partial \boldsymbol{Q}(\boldsymbol{\theta})}{\partial\theta_k}
\cdot \boldsymbol{Q}^{-1}(\boldsymbol{\theta})
\cdot \dfrac{\partial \boldsymbol{Q}(\boldsymbol{\theta})}{\partial\theta_l}
\right )
\end{equation}

Measurements at different frequencies, in general, contribute differently to the elements of the FIM. Thanks to the fact that in Cartesian and polar coordinates, the covariance matrix is purely and block diagonal, respectively, it is possible to define the contribution to the FIM element $(k,l)$ as a function of parameters $\boldsymbol{\theta}$ and frequency $\omega_i$:
\begin{equation}\label{eq:contribution}
\left[\Delta \boldsymbol{\mathcal{F}}_i(\boldsymbol{\theta})\right]_{k,l} = \dfrac{\partial z_i^\top(\boldsymbol{\theta})}{\partial\theta_k}\boldsymbol{Q}_i^{-1}(\boldsymbol{\theta})
\dfrac{\partial z_i(\boldsymbol{\theta})}{\partial\theta_l}
+\dfrac{1}{2}\text{tr}
\left(
\boldsymbol{Q}_i^{-1}(\boldsymbol{\theta})
\cdot\dfrac{\partial \boldsymbol{Q}_i(\boldsymbol{\theta})}{\partial\theta_k}
\cdot \boldsymbol{Q}_i^{-1}(\boldsymbol{\theta})
\cdot \dfrac{\partial \boldsymbol{Q}_i(\boldsymbol{\theta})}{\partial\theta_l}
\right ).
\end{equation}
To calculate $(k,l)$ element of $\boldsymbol{\mathcal{F}}$, we can use the following property.

\textbf{Property 1}. The element $(k,l)$ of the FIM can be expressed as a sum of contributions at every frequency $\omega_i$, $i = 1,\dots,N$:
\begin{equation}
    \left[ \boldsymbol{\mathcal{F}}(\boldsymbol{\theta},{\boldsymbol{\omega}})\right]_{k,l} = \sum_{i=1}^{N} \left[\Delta \boldsymbol{\mathcal{F}}_i(\boldsymbol{\theta})\right]_{k,l}.
    \label{pro:FIMdiagProperty}
\end{equation}

When expressed in Cartesian coordinates, the vector $z_i(\boldsymbol{\theta})$ from \eqref{eq:contribution} is $z_i(\boldsymbol{\theta}) = \left[R_i(\boldsymbol{\theta}),X_i(\boldsymbol{\theta})\right]^\top$
and $\boldsymbol{Q}_i(\boldsymbol{\theta})$ is a covariance matrix block:
\begin{equation}
\boldsymbol{Q}_i(\boldsymbol{\theta}) = 
\begin{bmatrix}
\sigma^2(R_i(\boldsymbol{\theta}))    & \sigma(R_i(\boldsymbol{\theta}),X_i(\boldsymbol{\theta}))\\ 
\sigma(R_i(\boldsymbol{\theta}),X_i(\boldsymbol{\theta}))  & \sigma^2(X_i(\boldsymbol{\theta}))
\end{bmatrix}=
\begin{bmatrix}
\alpha_i(\boldsymbol{\theta}) &  \beta_i(\boldsymbol{\theta})\\ 
\beta_i(\boldsymbol{\theta})  & \gamma_i(\boldsymbol{\theta})
\end{bmatrix},
\end{equation}
whose elements are expressed in terms of \textit{true} values of the impedance (magnitude $\rho_i$ and phase $\varphi_i$):
 \begin{align}
    \alpha_i(\boldsymbol{\theta})   &=     \rho_i^2  e^{-\sigma_\varphi^2}  \left[  \cos^2\varphi_i  (   \cosh(\sigma_\varphi^2) - 1  ) + \sin^2\varphi_i    \sinh(\sigma_\varphi^2  )  \right]\\
          &   +\sigma_{\rho_i}^2  e^{-\sigma_\varphi^2}  \left[  \cos^2\varphi_i     \cosh(\sigma_\varphi^2)       + \sin^2\varphi_i    \sinh(\sigma_\varphi^2  )  \right]\nonumber\\
    \beta_i(\boldsymbol{\theta})   &=     \rho_i^2  e^{-\sigma_\varphi^2}  \left[  \sin^2\varphi_i (   \cosh(\sigma_\varphi^2) - 1 ) + \cos^2\varphi_i    \sinh(\sigma_\varphi^2 )    \right]\\
          &   +\sigma_{\rho_i}^2  e^{-\sigma_\varphi^2}  \left[  \sin^2\varphi_i      \cosh(\sigma_\varphi^2)      + \cos^2\varphi_i     \sinh(\sigma_\varphi^2 )    \right]\nonumber\\
    \gamma_i(\boldsymbol{\theta})   &= \sin\varphi_i\cos\varphi_i  e^{-2\sigma_\varphi^2}  \left[  \sigma_{\rho_i}^2 + \rho_i^2  (1 - e^{\sigma_\varphi^2})  \right].
 \end{align}
Note that $\rho_i = \rho_i(\boldsymbol{\theta})$, $\varphi_i = \varphi_i(\boldsymbol{\theta})$, $\sigma_{\rho_i} =  \sigma_{\rho_i}(\boldsymbol{\theta})$ and $\sigma_\varphi = const.$

On the other hand, in polar coordinates $z_i(\boldsymbol{\theta}) = \left[\rho_i(\boldsymbol{\theta}), \varphi_i(\boldsymbol{\theta})\right]^\top$
and covariance matrix block is of a diagonal form:
\begin{equation}
    \boldsymbol{Q}_i(\boldsymbol{\theta})
    =
    \begin{bmatrix}
    \sigma_{\rho_i}^2(\boldsymbol{\theta})    & 0\\ 
    0  & \sigma_{\varphi}^2
    \end{bmatrix}.
\end{equation}

Using either the polar or Cartesian coordinates, the calculation of the CRLB requires the evaluation of the FIM elements at the true parameter values, $\boldsymbol{{\theta}_{\text{true}}}$, which are unknown in practice:
\begin{equation}\label{eq:CRLB_evaluation}
\small
[ \boldsymbol{\mathcal{F}}]_{k,l} = [ \boldsymbol{\mathcal{F}}(\boldsymbol{\theta}, \boldsymbol{\omega})]_{k,l}
\Big|_{\scriptsize\begin{matrix}
        \boldsymbol{\theta} = \boldsymbol{{\theta}_{\text{true}}}
        \end{matrix}}.
\end{equation}
The diagonal elements of the FIM inverse, $\boldsymbol{\mathcal{C}} \triangleq \boldsymbol{\mathcal{F}}^{-1}$, evaluated at parameters' true values and frequency values, provide CRLB for each parameter and define the least possible variance of the estimated parameter. Namely:
\begin{equation}
   \sigma^2_{\theta_i} \triangleq \text{Var}(\theta_i) \geqslant \boldsymbol{\mathcal{C}}_{i,i},\; \forall i = 1,\dots, M.
\end{equation}

\pagebreak
\subsection{Least-variance Circuit Parameter Identification via CRLB}
When performing EIS measurements, it is common to use logarithmically spaced frequency points, say $N$, from a minimum $f_{min}$ to a maximum $f_{max}$ desired frequency to measure the cell's impedance. 

As known from classical information theory, we expect that different circuit parameters are dominant at different frequency ranges, i.e. different parameters achieve a maximum contribution to the FIM at different frequencies (see Fig.~\ref{fig:threeSubPlotsContrib} of Section \ref{sec:Contributions}). Therefore, for a fixed single frequency point, each parameter contributes differently to the FIM. Modifying such frequency affects the variances of all estimated parameters and justifies the search for frequencies that improve the overall estimation accuracy.

For the multi-parameter estimation problem where all the parameters are mutually coupled, there exist different techniques one can use to improve the FIM and, consequently, minimize the variances of the estimates. Some of these include maximizing: the trace of FIM, the determinant of FIM, and the smallest eigenvalue of FIM, known as A-, D- and E-optimal experiment design, respectively.

In this section, we propose an algorithm to find a set of frequencies that provide measurements carrying the most information to estimate ECM parameters by maximizing the smallest eigenvalue of the FIM.

\subsection{E-optimal design}

The square FIM, $ \boldsymbol{\mathcal{F}}$, is a real, symmetric, positive-semidefinite matrix. Therefore, it has $M$ real and non-negative eigenvalues.
Let's define the $M$-dimensional vector of the FIM's eigenvalues as 
\begin{equation}
    \Lambda= \left[\lambda_1,\dots,\lambda_M\right]^\top \in\mathbb{R}^M
\end{equation}
with eigenvalues in non-descending order $0\leqslant\lambda_1 \leqslant \dots \leqslant \lambda_M$.

The eigenvalues of the FIM's inverse $\boldsymbol{\mathcal{C}}$ are equal to the reciprocal values of the eigenvalues of the FIM $ \boldsymbol{\mathcal{F}}$. Therefore, the eigenvalues of $\boldsymbol{\mathcal{C}}$ are
\begin{equation}
    \dfrac{1}{\lambda_1},\dots,\dfrac{1}{\lambda_M}\; \text{with}\; \dfrac{1}{\lambda_1} \geqslant \dots \geqslant\dfrac{1}{\lambda_M} \geqslant 0.
\end{equation}

As known, square roots of eigenvalues of FIM's inverse, $\boldsymbol{\mathcal{C}} = \boldsymbol{\mathcal{F}}^{-1}$, represent the length of semi-axes of estimated parameters' confidence ellipsoid \cite{nielsen_elementary_2018}. 
Since the smallest eigenvalue of $ \boldsymbol{\mathcal{F}}$ defines the highest eigenvalue of $\boldsymbol{\mathcal{C}}$, to minimize the length $\sqrt{1/\lambda_1}$ of the largest semi-axis of the confidence ellipsoid, it is sufficient to maximize the smallest eigenvalue of the $ \boldsymbol{\mathcal{F}}$ matrix, $\lambda_1$. We will show that it consequently results in a decrease of the volume of $M-$dimensional confidence ellipsoid given by:
\begin{equation}
    V_M = \dfrac{2}{M}
    \dfrac{\pi^{\tfrac{M}{2}}}{\Gamma\left(\tfrac{M}{2}\right)}
    \prod_{i=1}^M \sqrt{1/\lambda_i},
    \label{eq:ellipsoid_volume}
\end{equation}
where $\Gamma$ is the gamma function \cite{wilson_volume_2009}. It is worth mentioning that we neglect the exact correspondence between eigenvalues and parameters but focus on minimizing the largest axis of the confidence ellipsoid to improve the overall accuracy.

\subsection{Algorithm: Frequencies Adjustments via E-optimal Design}
The proposed algorithm's objective is to adjust the initial set of frequencies so that the new frequency set provides better estimates than the traditional log-spaced frequency span. The initial and final set of frequencies will have the same predetermined number of points, $N$ and the frequency range $\left[f_{min},f_{max}\right]$. With $\Omega = \left\{1,2,\dots,N\right\}$, we denote the set of indices of frequency measurement points.
\newcommand{\Break}{\State \textbf{break} }

\begin{algorithm}[H]
\caption{Frequencies Adjustments via E-optimal Design}\label{alg:eigvalues_freq_adjustment}
\begin{algorithmic}[1]
\State \textbf{Initialize} $k = 0$, $\Omega = \{1,\dots,N\}$, $\Omega_c = \O$, $f_{min}$, $f_{max}$, $\mu$ \label{lst:Initialization}
  \Do
        \If{$k = 0$} \label{lst:ZmeasureStart}
            \State $\boldsymbol{\omega}^{k} = \texttt{logspace}(2\pi f_{min}, 2\pi f_{max}, N)$ \label{lst:PrelimIterStart}
            \State Obtain $\tilde{Z}(\boldsymbol{\omega}^{k}) = \left[\tilde{Z}(\omega_1^{k}),\dots,\tilde{Z}(\omega_N^{k})\right]^\top$
            \State $\boldsymbol{\theta}_{init} \gets \boldsymbol{\theta}_{0}$ (from the procedure described in Sec. \ref{sec:ECM_initial}) \label{lst:PrelimIterEnd}
        \Else
            \State $\boldsymbol{\omega}^{k} \gets \boldsymbol{\omega}^{k-1}$
            \State Obtain only $\tilde{Z}(\omega_m^{k})$
            \State Update: $\tilde{Z}(\boldsymbol{\omega}^{k}) \gets \left[\tilde{Z}(\omega_1^{k-1}),\dots,\tilde{Z}(\omega_m^{k}),\dots,\tilde{Z}(\omega_N^{k-1})\right]^\top$
            \State $\boldsymbol{\theta}_{init} \gets \hat{\boldsymbol{\theta}}^{k-1}$
        \EndIf \label{lst:ZmeasureEnd}
        \State Estimate the ECM parameters, $\hat{\boldsymbol{\theta}}^k$ by solving \eqref{eq:objective_fun} with initial guess $\boldsymbol{\theta}_{init}$ \label{lst:EstimStart}
        \State Compute FIM: $ \boldsymbol{\mathcal{F}}^k = \texttt{FIM}(\boldsymbol{\omega}^k,\hat{\boldsymbol{\theta}})$ using \eqref{eq:CRLB_exp}
        \State Compute eigenvalues of FIM: $\Lambda^k = \texttt{eig}( \boldsymbol{\mathcal{F}}^k) = \left[ \lambda_1^k,\dots,\lambda_M^k \right]^\top$ \label{lst:EstimEnd}
        \For{$i\in\Omega\setminus\Omega_c$} \label{lst:DecidingAdjStart}
            \State $\Delta \omega^k_i = \frac{\omega^k_i}{\mu} $
            \If{$i = N$}
                \State $\omega_i^k \gets \omega_i^k-\Delta \omega^k_i$
            \Else
                \State $\omega_i^k \gets \omega_i^k+\Delta \omega^k_i$
            \EndIf
            \State $ \boldsymbol{\mathcal{F}}^k_{\Delta i} = \texttt{FIM}(\boldsymbol{\omega}^k, \hat{\boldsymbol{\theta}})$
            \State $\Lambda_{\Delta i}^k = \texttt{eig}( \boldsymbol{\mathcal{F}}^k_{\Delta i})$
            \State $d_i^k = \dfrac{\min\Lambda^k-\min\Lambda_{\Delta i}^k}{\Delta \omega^k_i}$
        \EndFor
        \State $d_m^k = {\argmax}_{d_i^k}\{|d_i^k|: i\in\Omega\setminus\Omega_c\}$
        \State $s = \sgn(d_m^k)$ \label{lst:DecidingAdjEnd}
        \State $n \gets 1$ \label{lst:FreqAdjustStart}
        \Do
            % \State $\boldsymbol{\omega}^k_{n\Delta m} = \left[\omega_1^k,\dots,\omega_m^k+ s\cdot n\Delta \omega^k_m,\dots,\omega_N^k\right]^\top$
            % \State ${\omega}^k_m \gets \omega_m^k+ s\cdot n\Delta \omega^k_m$
            \State $ {\omega}^k_m \gets \omega_m^k+ s\cdot n\Delta \omega^k_m $
                \If{${\omega}^k_m \geqslant 2\pi f_{max}$ \text{\textbf{or}} ${\omega}^k_m \leqslant 2\pi f_{min}$} \label{lst:fminfmaxStart}                    
                \State ${\omega}^k_m \gets 2\pi\left[(s-1)f_{min} + (1-s)f_{max}\right]$ 
                    \Break
                \EndIf \label{lst:fminfmaxEnd}
            \State $ \boldsymbol{\mathcal{F}}^k_{n\Delta m} = \texttt{FIM}(\boldsymbol{\omega}^k,\hat{\boldsymbol{\theta}})$
            \State $\Lambda_{n\Delta m}^k = \texttt{eig}( \boldsymbol{\mathcal{F}}^k_{n\Delta m})$
            \State $n\gets n+1$
        \doWhile{$\min\Lambda_{n\Delta m}^k > \min\Lambda_{(n-1)\Delta m}^k$}
        % \State $\boldsymbol{\omega}^k \gets \left[\omega_1^k,\dots,\omega_m^k+ (n-1)\Delta \omega^k_m,\dots,\omega_N^k\right]^\top$
        \State $\omega_m^k \gets \omega_m^k + s\cdot (n-1)\Delta\omega_m^k$ 
        \State $\Omega_c \gets \Omega_c \cup \{m\}$ \label{lst:FreqAdjustEnd}
        \State $k\gets k+1$
  \doWhile{$\Omega \neq \O$} % <--- use \doWhile for the "while" at the end
\end{algorithmic}
\end{algorithm}

Globally, the algorithm's iterations consist of the following main parts: perform the EIS measurements at defined frequencies (lines~\ref{lst:ZmeasureStart}-\ref{lst:ZmeasureEnd}), estimating the ECM parameters, FIM and its eigenvalues (lines~\ref{lst:EstimStart}-\ref{lst:EstimEnd}), decide which frequency to adjust (lines~\ref{lst:DecidingAdjStart}-\ref{lst:DecidingAdjEnd}) and optimally adjust it using a gradient approach (lines~\ref{lst:FreqAdjustStart}-\ref{lst:FreqAdjustEnd}). The algorithm terminates when adjustments of all the frequencies are identified and implemented.

The first stage (line~\ref{lst:Initialization}) initializes the iteration counter $k$, the set of frequency measurements' indices $\Omega$ and the set of already adjusted and fixed frequencies, $\Omega_c$. As input, the user defines the minimum and maximum frequencies for the EIS measurements, $f_{min}$ and $f_{max}$, and the parameter $\mu$, used to numerically compute small perturbations of the frequencies through the iterations. 

In the preliminary iteration (for $k = 0$) the modeller can use $N$ frequency points spread logarithmically between defined frequency bounds and obtain $N$ impedance measurements. ECM parameters can be initialized by applying the described procedure in Section \ref{sec:ECM_initial}, $\boldsymbol{\theta}_{init} = \boldsymbol{\theta}_0$. In all the other iterations, the impedance is measured only at the frequency adjusted in the previous iteration. ECM parameters take the initial values equal to the estimated ones from the previous iteration, $\boldsymbol{\theta}_{init} = \hat{\boldsymbol{\theta}}^{k-1}$. On estimating the ECM parameters, we compute the FIM calculated using the estimated parameters' values and find its eigenvalues using the procedures $\texttt{FIM}$ and $\texttt{eig}$.

The lines~\ref{lst:DecidingAdjStart}-\ref{lst:DecidingAdjEnd} of the pseudo-code decide which frequency to adjust (from the set of frequencies that have not been adjusted). It finds the index of the frequency for which adjustment has the most impact on the minimum eigenvalue of the FIM. This step is done numerically: the algorithm perturbs each candidate frequency $\omega_i$, $i\in\Omega\setminus\Omega_c$, one by one, by increasing it by its small fraction, $\Delta\omega_i$. To satisfy the pre-determined frequency bounds $f_{min}$ and $f_{max}$, the highest frequency point, $\omega_N$ is decreased by $\Delta\omega_i$ and we assume that $\mu$ is chosen such that for all the other frequencies, $\Delta\omega_i$ is sufficiently small so that the perturbed frequency stays within the bounds. In that case, one should ensure:
\begin{equation}
\omega_{N-1} + \Delta\omega_{N-1} =  \omega_{N-1} + \dfrac{\omega_{N-1}}{\mu} \leqslant 2\pi f_{max},  
\end{equation} 
and therefore 
\begin{equation}
    \mu \geqslant \dfrac{\omega_{N-1}}{2\pi f_{max} - \omega_{N-1}}.
\end{equation}
The FIM and its eigenvalues are re-computed for each perturbation. In $k$-th iteration, the metric used to quantify the impact of the perturbation of $\omega_i$ on the minimum eigenvalue of the FIM, is denoted as $d_{i}^k$. It represents a sensitivity of the minimum eigenvalue of the FIM to the perturbation of frequency $\omega_i$. By finding the maximum sensitivity (by the absolute value, since $d_i^k$ can also take negative values), the algorithm decides the index of the best suitable frequency to adjust and the direction of the adjustment (i.e., depending on the sign of the sensitivity).

In lines \ref{lst:FreqAdjustStart}-\ref{lst:FreqAdjustEnd} of the pseudo-code, the algorithm further perturbs the chosen frequency (index denoted as $m$) in discreet steps by multiplying $\Delta\omega_m^k$ by the loop counter $n$ and the sensitivity sign $s$. In case the frequency bounds are violated, by increasing or decreasing the frequency, the algorithm fixes the current frequency to the maximum or minimum frequency, respectively (lines~\ref{lst:fminfmaxStart}-\ref{lst:fminfmaxEnd}). It then re-computes the FIM and its eigenvalues until there is no further improvement (increase) of the minimum eigenvalue. Therefore, the frequency point is considered to be adjusted and fixed. Its index is then included in the set of already adjusted frequencies $\Omega_c$.

\pagebreak
\section{Results and Discussion}
In this section, we present the results of numerical simulations used to verify the proposed methods for parameters initialization, estimation, accuracy assessment and frequencies adjustment.
\subsection{Cell and Measurement Model}

\textbf{Circuit Topology.}
We use the ECM for the Li-ion battery cell shown in Fig. \ref{fig:SpectrumAndECM}. The parameter vector is:
\begin{equation}
    \boldsymbol{\theta} = \left[R_s, Q_{HF}, \phi_{HF}, R_1, Q_1, \phi_1, R_2, Q_2, \phi_2, Q_W\right]^\top.
\end{equation}
For simulation purposes, we assume to know the ECM parameters' \textit{true values}, listed in Tab.~\ref{tab:ECMparamValues}.

\noindent\textbf{Impedance Data Generation.}
Artificial EIS measurements are generated by assuming that the topology and true values of the Full Randles ECM parameters are known and equal to $\boldsymbol{\theta}_\text{true}$. The measurement errors follow Gaussian distribution according to the discussed measurement model from Sec. \ref{sec:measurementModel}, with relative error in magnitude $\varepsilon_\rho = 1\%$ and absolute error in phase $\varepsilon_{\varphi} = 1^\circ$ (these values are quite typical for commercial EIS spectrometers). Therefore, here we assumed that our EIS measurements belong within a unique accuracy contour, let's say $A_1$. The following pseudo-code describes the generation of noisy impedance measurements within a specific accuracy contour plot.
\begin{algorithm}[H]
\caption{EIS Data Generation}\label{alg:dataGen}
\begin{algorithmic}[1]
\State \textbf{Initialize} $\boldsymbol{\omega}$, $\boldsymbol{\theta}_\text{true}$, ECM topology, $\varepsilon_\rho$, $\varepsilon_\varphi$
\For{$i = 1:N$}
    \State $R_i = R_i(\boldsymbol{\theta}_\text{true})$, $X_i = X_i(\boldsymbol{\theta}_\text{true})$
    \State $Z_i = R_i + j X_i$
    \State $\rho_i = |Z_i|$, $\varphi_i = \arg(Z_i)$
    \State $\Delta\rho \gets \mathcal{N}(0,\frac{\rho_i  \varepsilon_{\rho}}{3} )$
    \State $\Delta\varphi \gets \mathcal{N}(0,\frac{\varepsilon_{\varphi}}{3})$
    \State $\tilde{\rho}_i = \rho_i + \Delta\rho$
    \State $\tilde{\varphi}_i = \varphi_i + \Delta\varphi$
    \State $\tilde{Z}_i = \tilde{\rho}_i e^{j\tilde{\varphi}_i}$
    \State $\tilde{R}_i = \Re\{\tilde{Z}_i\}$, $\tilde{X}_i = \Im\{\tilde{Z}_i\}$
\EndFor
\end{algorithmic}
\end{algorithm}
It is important to mention that, according to the EIS instrument characteristics, the normal distribution of measurement errors is valid in polar coordinates for both magnitude and phase. After the projection from polar to Cartesian coordinates, residuals for the real and imaginary parts are, in general, no longer normally distributed. However, for considered values of  $\varepsilon_\rho = 1\%$ and $\varepsilon_{\varphi} = 1^\circ$, after the projection, errors of the real and imaginary part are practically normal \cite{milano_static_2016}. This is confirmed by comparing the quantile-quantile (QQ) plots for the noise of both real and imaginary parts of simulated impedance measurements (of a corresponding error structure) with the ones of a standard normal random variable. An interested reader can generate these QQ plots and easily verify this statement.

\subsection{Parameters' Initial Guess and Fitted Values Accuracy}

In what follows, we analyze the accuracy of the parameters' initial guess and fitted values after carrying out a controlled numerical experiment with known true values of the parameters. To show that the fitting algorithm provides the estimated values of the parameters close to the true values with an expected accuracy, we performed 1000 simulations. In each simulation, we generate the noisy EIS data as described in Algorithm~\ref{alg:dataGen} for a logarithmically spread set of frequencies, from $f_{min} = 10^{-2}$ Hz to $f_{max} = 10^4$ Hz and 10 points per decade. All the ECM parameters are first initialized by applying the described process of Section \ref{sec:InitializationLF_HF_MF}  and then estimated by solving the optimization problem \eqref{eq:objective_fun_pol}, in polar, and \eqref{eq:objective_fun_cart}, in Cartesian coordinates. Tab. \ref{tab:true_and_initial_params} shows the mean values of the parameters' initial guess after simulations. The proposed method for the initialization, where all ten parameters are initialized only by applying the properties of the EIS spectrum and before running the WCNLS fitting method, provides approximate values which do not exceed $16\%$ for $R_s$, while most of the parameters are initialized with the accuracy better than $8\%$.

\begin{table}[H]
\centering
\begin{tabular}{cccc}
\hline
\multicolumn{1}{|c|}{Param.} & \multicolumn{1}{c|}{$\theta$} & \multicolumn{1}{c|}{$\mu_{\theta_0}$} & \multicolumn{1}{c|}{$\left|\frac{\theta-\theta_0}{\theta}\right|\cdot 100$ (\%)} \\ \hline\hline
\multicolumn{1}{|c|}{$R_s$} & \multicolumn{1}{c|}{3.800E-02} & \multicolumn{1}{c|}{4.227E-02} & \multicolumn{1}{c|}{15.90} \\ \hline
\multicolumn{1}{|c|}{$Q_{HF}$} & \multicolumn{1}{c|}{1.667E+04} & \multicolumn{1}{c|}{1.736E+04} & \multicolumn{1}{c|}{7.53} \\ \hline
\multicolumn{1}{|c|}{$\varphi_{HF}$} & \multicolumn{1}{c|}{-8.500E-01} & \multicolumn{1}{c|}{-8.519E-01} & \multicolumn{1}{c|}{0.70} \\ \hline
\multicolumn{1}{|c|}{$R_1$} & \multicolumn{1}{c|}{4.500E-01} & \multicolumn{1}{c|}{4.714E-01} & \multicolumn{1}{c|}{4.75} \\ \hline
\multicolumn{1}{|c|}{$Q_1$} & \multicolumn{1}{c|}{2.000E-02} & \multicolumn{1}{c|}{1.815E-02} & \multicolumn{1}{c|}{10.44} \\ \hline
\multicolumn{1}{|c|}{$\varphi_1$} & \multicolumn{1}{c|}{9.000E-01} & \multicolumn{1}{c|}{9.425E-01} & \multicolumn{1}{c|}{4.72} \\ \hline
\multicolumn{1}{|c|}{$R_2$} & \multicolumn{1}{c|}{6.500E-01} & \multicolumn{1}{c|}{6.382E-01} & \multicolumn{1}{c|}{1.90} \\ \hline
\multicolumn{1}{|c|}{$Q_2$} & \multicolumn{1}{c|}{4.000E-01} & \multicolumn{1}{c|}{3.593E-01} & \multicolumn{1}{c|}{12.38} \\ \hline
\multicolumn{1}{|c|}{$\varphi_2$} & \multicolumn{1}{c|}{9.000E-01} & \multicolumn{1}{c|}{9.425E-01} & \multicolumn{1}{c|}{4.72} \\ \hline
\multicolumn{1}{|c|}{$Q_W$} & \multicolumn{1}{c|}{3.693E+00} & \multicolumn{1}{c|}{3.537E+00} & \multicolumn{1}{c|}{4.24} \\ \hline
\end{tabular}
\caption{True and initial ECM parameters' values.}
\label{tab:true_and_initial_params}
\end{table}

The CNLS estimation results are presented in Tab.~\ref{tab:true_and_estimated_params}. Both formulations in polar and Cartesian coordinates provide the same results. The mean value $\mu_{\hat{\theta}}$ of each estimated parameter is extremely close to its true value (maximum relative error of $1.134\%$ for $Q_{HF}$). Therefore, our estimator is unbiased. We calculated the variance of each parameter from the collection of all the estimated parameters' values. On the other hand, by having access to the parameters' true values, we computed the CRLB, defining the best possible variance of estimated parameters for an unbiased estimator. The calculated variance from 1000 simulations is quite close to the exact CRLB for the first nine parameters. This variance is slightly lower for the $Q_W$, due to the finite number of repeated simulations. However, the results indicate that our estimator is efficient, i.e., the estimator is capable of attaining the CRLB.

\begin{table}[H]
\begin{tabular}{|c|c|c|c|c|c|c|c|}
\hline
Param. & $\theta$ & $\mu_{\hat{\theta}}$ & $\left|\frac{\theta-\mu_{\hat{\theta}}}{\theta}\right|$ & $\sigma^2_\theta$ & $CRLB_{\theta}$ & $\dfrac{\sigma^2_\theta}{\theta^2}$ & $\dfrac{CRLB_\theta}{\theta^2}$ \\ \hline\hline
$R_s$ & 3.800E-02 & 3.801E-02 & 0.752\% & 1.279E-07 & 1.159E-07 & 8.860E-03 & 8.028E-03 \\ \hline
$Q_{HF}$ & 1.667E+04 & 1.668E+04 & 1.134\% & 5.602E+04 & 5.065E+04 & 2.017E-02 & 1.823E-02 \\ \hline
$\varphi_{HF}$ & -8.500E-01 & -8.500E-01 & 0.129\% & 1.902E-06 & 1.723E-06 & 2.633E-04 & 2.385E-04 \\ \hline
$R_1$ & 4.500E-01 & 4.499E-01 & 0.476\% & 7.220E-06 & 6.860E-06 & 3.565E-03 & 3.388E-03 \\ \hline
$Q_1$ & 2.000E-02 & 1.999E-02 & 0.956\% & 5.743E-08 & 5.335E-08 & 1.436E-02 & 1.334E-02 \\ \hline
$\varphi_1$ & 9.000E-01 & 9.001E-01 & 0.201\% & 5.075E-06 & 4.666E-06 & 6.266E-04 & 5.761E-04 \\ \hline
$R_2$ & 6.500E-01 & 6.502E-01 & 0.652\% & 2.806E-05 & 2.788E-05 & 6.641E-03 & 6.599E-03 \\ \hline
$Q_2$ & 4.000E-01 & 4.000E-01 & 0.585\% & 8.788E-06 & 8.710E-06 & 5.492E-03 & 5.444E-03 \\ \hline
$\varphi_2$ & 9.000E-01 & 8.999E-01 & 0.484\% & 2.965E-05 & 2.921E-05 & 3.660E-03 & 3.606E-03 \\ \hline
$Q_W$ & 3.693E+00 & 3.693E+00 & 0.464\% & 4.560E-04 & 4.586E-04 & 3.344E-03 & 3.363E-03 \\ \hline
\end{tabular}
\caption{True and estimated ECM parameters' values and the estimation accuracy.}
\label{tab:true_and_estimated_params}
\end{table}

\begin{figure}[H]
  \begin{subfigure}[t]{.45\textwidth}
    \centering
    \includegraphics[scale = 0.38]{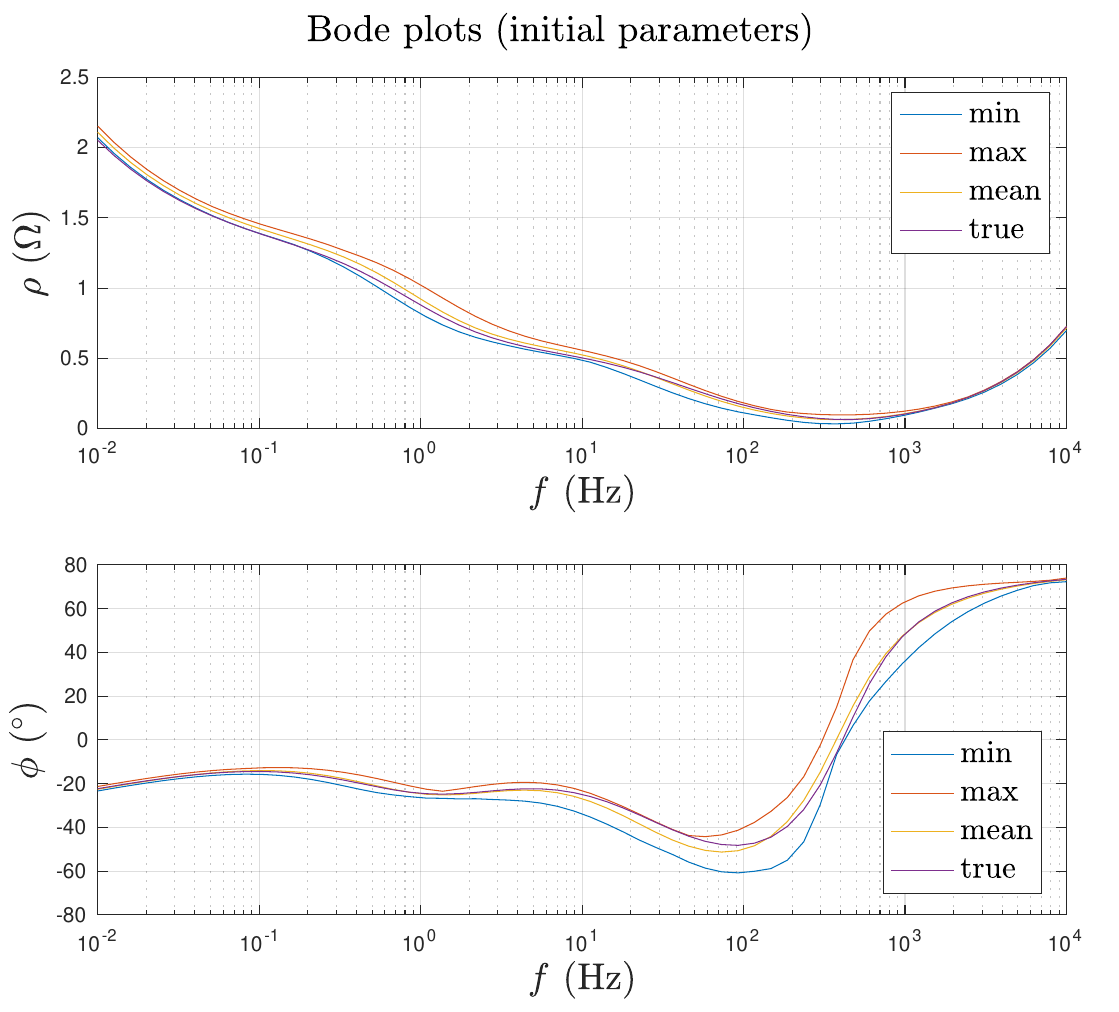}
    \caption{Initial values of ECM parameters.}
    \label{fig:a_Bode_initialized}
  \end{subfigure}
  \hfill
  \begin{subfigure}[t]{.45\textwidth}
    \centering
    \includegraphics[scale = 0.38]{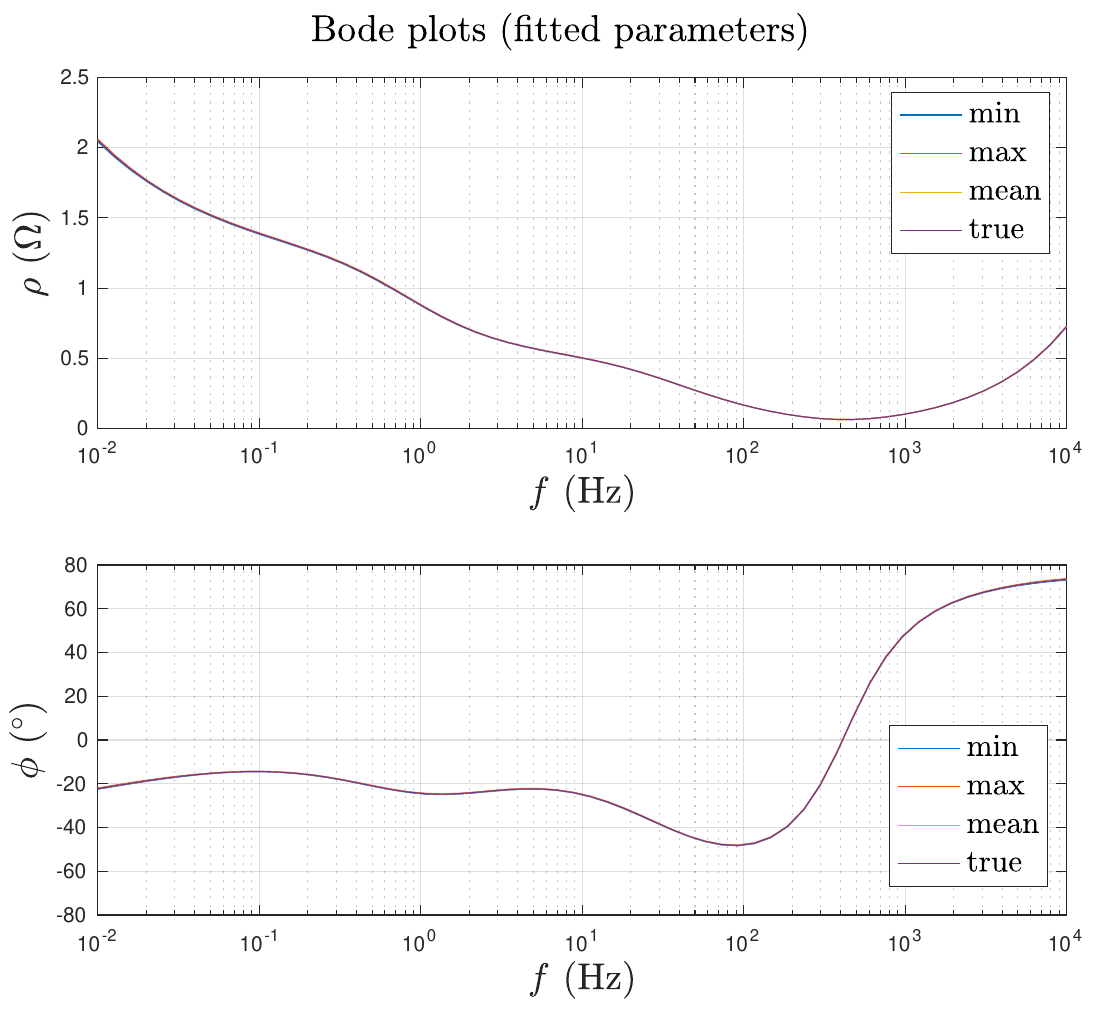}
    \caption{Fitted values of ECM parameters.}
    \label{fig:b_Bode_fitted}
    
  \end{subfigure}

  \medskip

  \begin{subfigure}[t]{.45\textwidth}
    \centering
    \includegraphics[scale = 0.38]{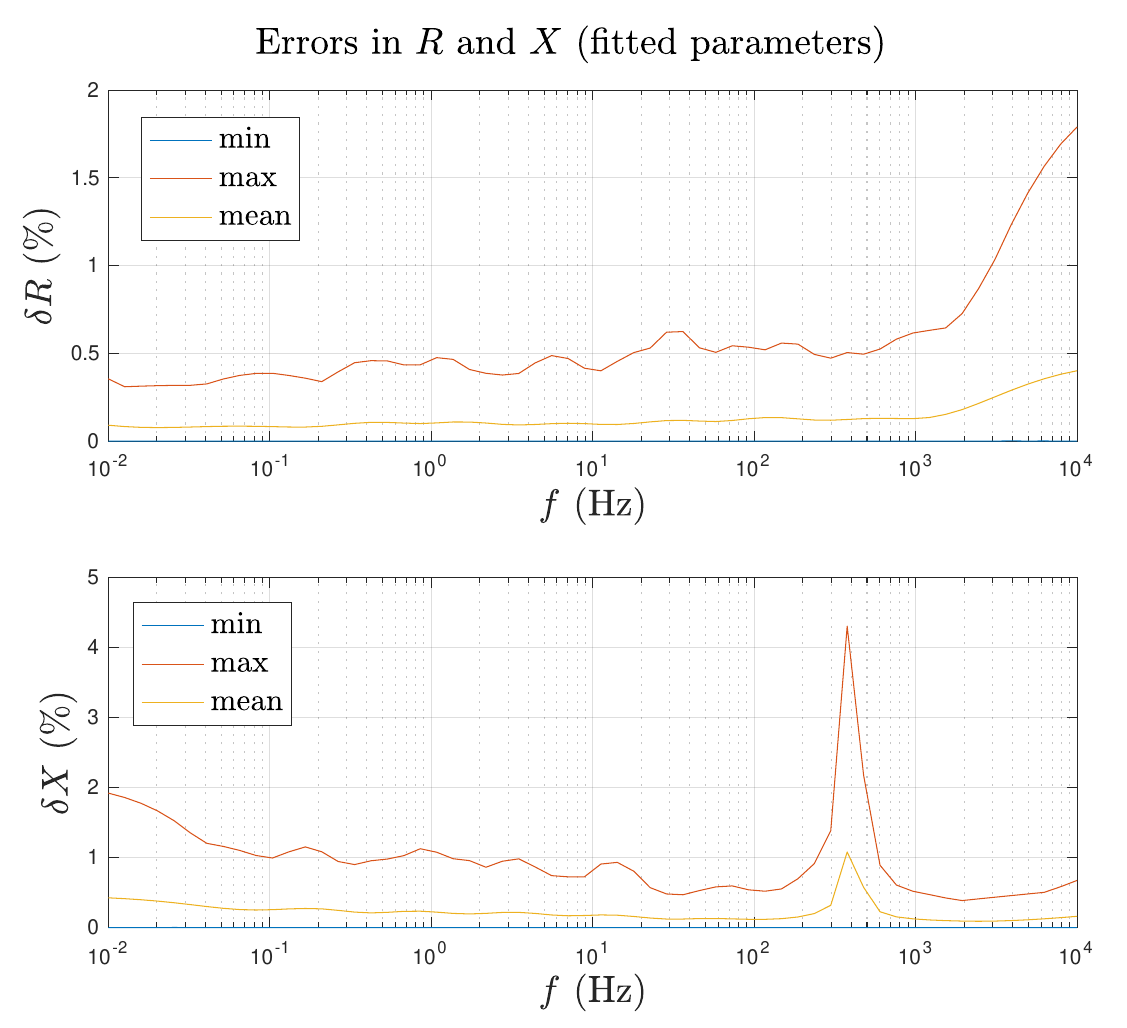}
    \caption{Relative errors in $R$ and $X$.}
    \label{fig:c_R_X_errors_fitted}
  \end{subfigure}
  \hfill
  \begin{subfigure}[t]{.45\textwidth}
    \centering
    \includegraphics[scale = 0.38]{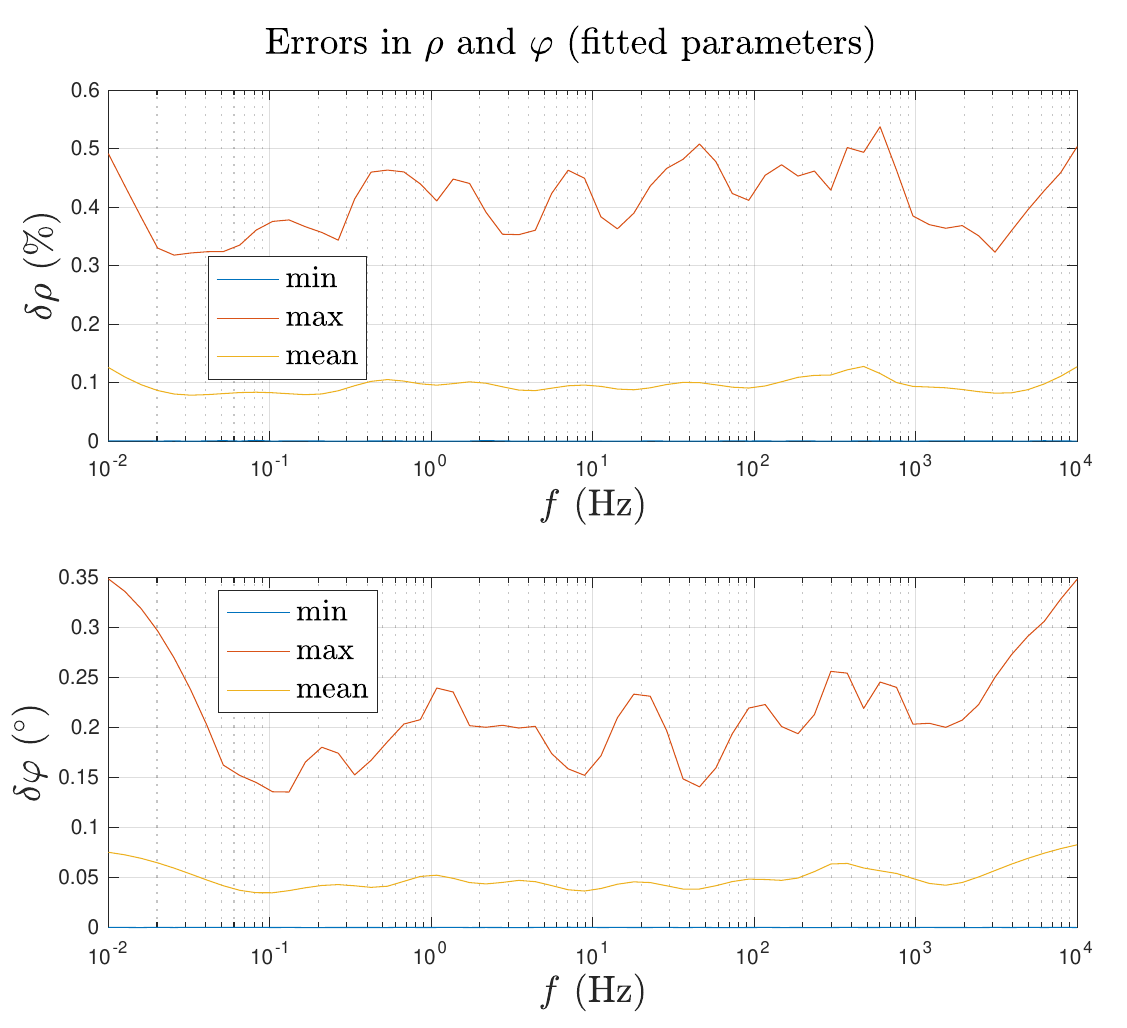}
    \caption{Relative error in $\rho$ and absolute error in $\varphi$.}
    \label{fig:d_rho_phi_errors_fitted}
  \end{subfigure}
  \caption{Bode plots obtained using the initial and estimated parameters (a-b) and minimum, maximum and mean errors of reconstructed EIS spectra compared to the true values (c-d).}
\end{figure}

To illustrate the accuracy of the initialized and estimated parameters' values, in each simulation, the EIS data is reconstructed by evaluating the equivalent impedance functions given by \eqref{eq:zeq_real} and \eqref{eq:zeq_imag} at the obtained initial and estimated value for the defined frequency span. From the entire families of Bode curves, in Fig. \ref{fig:a_Bode_initialized} and Fig. \ref{fig:b_Bode_fitted}, we show the minimum, maximum, mean, and true value of the impedance magnitude and phase at a specific frequency, using the initial and fitted estimated parameters' values, respectively. Since the curves in Fig. \ref{fig:b_Bode_fitted} are extremely close, Fig. \ref{fig:c_R_X_errors_fitted} and \ref{fig:d_rho_phi_errors_fitted} show minimum, maximum and mean values of the errors in Cartesian and polar coordinates, respectively, from the reconstructed EIS data using the estimated values, with reference to the true EIS spectrum.

\pagebreak
\subsection{Parameters' Contributions to the FIM}\label{sec:Contributions}
\begin{figure}[H]
    \centering
    \includegraphics[scale = 0.55]{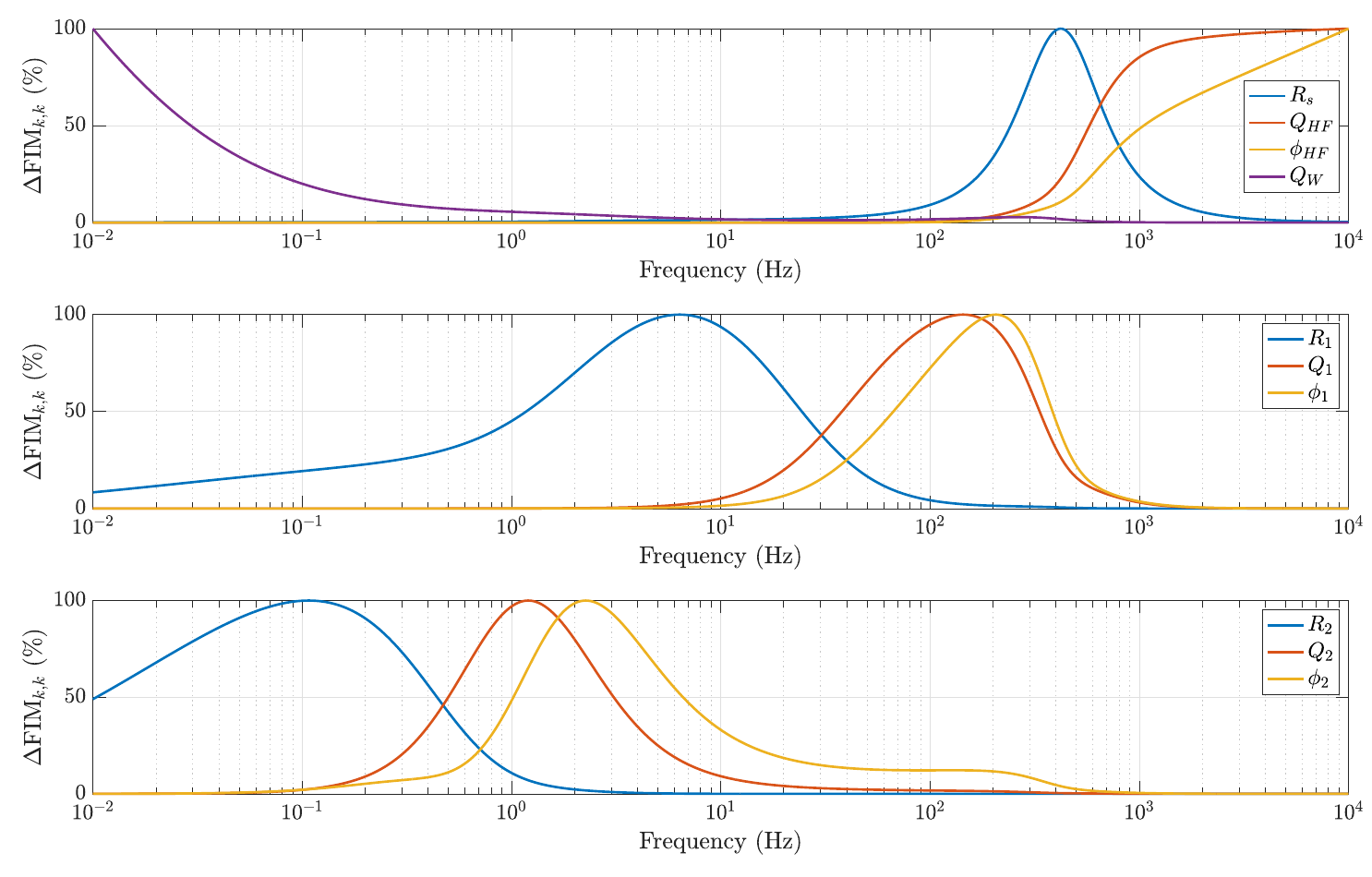}
    \caption{Contributions to the diagonal elements of FIM as a function of frequencies for every ECM parameter.}
    \label{fig:threeSubPlotsContrib}
\end{figure}

Using the expression \eqref{eq:contribution}, for a certain frequency region, it is possible to show the evolution of different contributions to the FIM, as a function of frequency. In Fig. \ref{fig:threeSubPlotsContrib}, the contributions to the diagonal elements of FIM are shown for the frequency range from $f_{min} = 10^{-2}$ Hz to $f_{max} = 10^4$ Hz. For visibility, each curve is normalized by its peak value achieved in this range.

For different parameters, different frequencies provide measurements at which the diagonal elements of the FIM reach their peak values. As expected, contributions from the elements $R_s$ and $Q_{HF}$ and $\phi_{HF}$ are the highest at high frequencies, while from $Q_W$ is increasing towards low frequencies. Peaks of the contribution curves corresponding to CPE1 parameters ($R_1, Q_1, \phi_1$) and CPE2 ($R_2, Q_2, \phi_2$) are within the mid-frequency region, in the expected order, having the time constants $\tau_1$ and $\tau_2$ where $\tau_1 < \tau_2$.

\subsection{Parameters and Variances Identification}
In this part, we present the results obtained after executing the proposed algorithm for frequency adjustment. The purpose of the numerical study is to compare the performance of the proposed algorithm for frequency adjustments and show that for that measurements at resulting frequencies convey more information than ones performed at log-spaced frequencies. Therefore the overall accuracy of parameters' estimated values increases.

\begin{figure}[htb!]
    \centering
        \includegraphics[width=\textwidth]{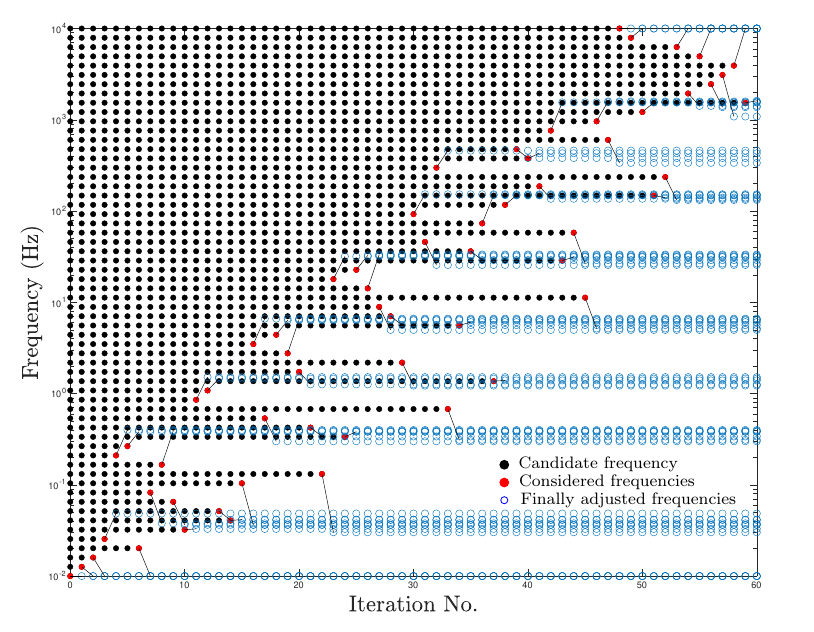}
    \caption{Frequencies adjustments via E-optimal design.}
    \label{fig:EIGalgIllustrative}
\end{figure}

Fig. \ref{fig:EIGalgIllustrative} illustrates the adjustment of frequencies throughout the iterations. Frequencies with $i\in\Omega$ in the current iteration are shown in black dots. The frequency point about to be adjusted (as a result of lines \ref{lst:DecidingAdjStart}-\ref{lst:DecidingAdjEnd} of the pseudo-code) are indicated in red. After the frequency is finally adjusted and fixed, it is indicated as a blue empty dot connected to the value in the previous iteration to track the change.

\begin{figure}[H]
    \centering
    \includegraphics[scale = 0.6]{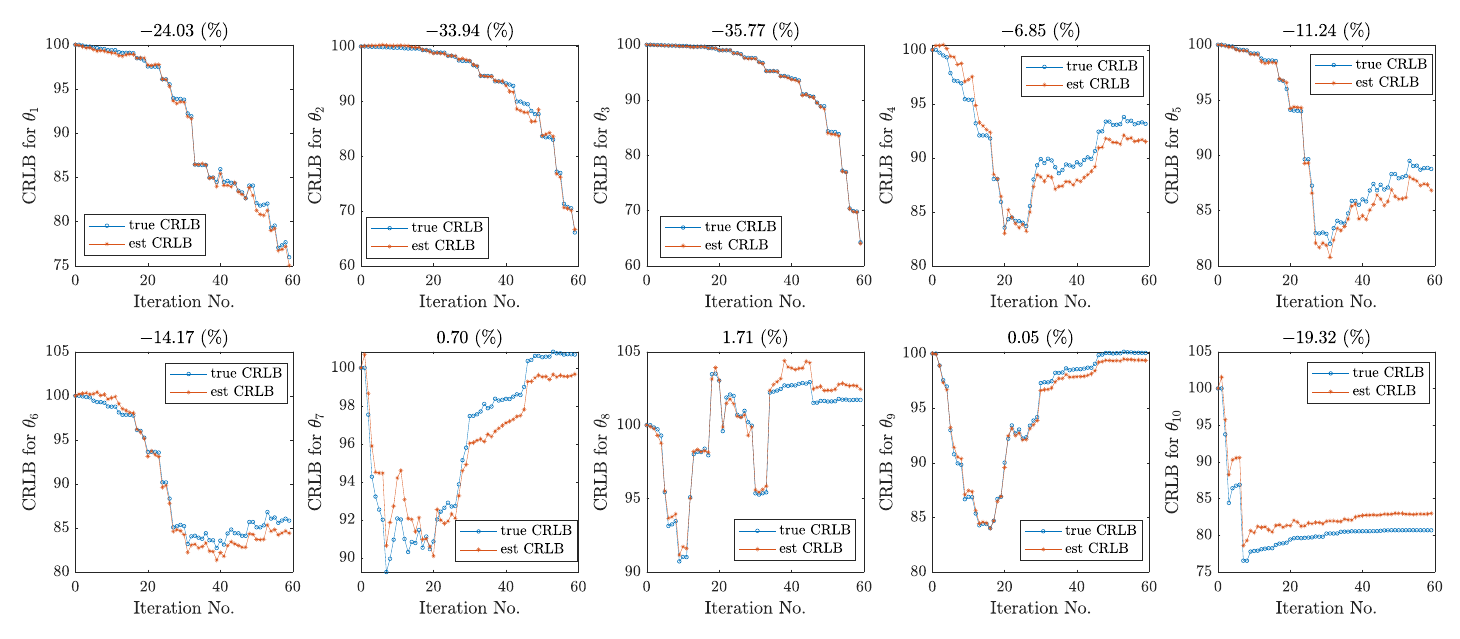}
    \caption{Relative improvement in CRLB for each ECM parameter compared to the initial log-spaced frequency span.}
    \label{fig:CRLB_tracking}
\end{figure}

At each iteration, we compare true CRLB, calculated using the parameters' true values, $\boldsymbol{\theta}$ with its approximate value, which is evaluated at the parameters’ estimated values, at the current iteration, ${\boldsymbol{\hat{\theta}}}$.
Fig.~\ref{fig:CRLB_tracking} shows the evolution of the CRLB throughout the iterations, namely, $\boldsymbol{\mathcal{C}}_{i,i}(\boldsymbol{\theta}, \boldsymbol{\omega}^k)$ and $\boldsymbol{\mathcal{C}}_{i,i}({\boldsymbol{\hat{\theta}}}^k, \boldsymbol{\omega}^k)$, $i = 1,\dots,M$ in a percentage of the CRLB at the log-spaced frequencies at the beginning of the algorithm. The CRLB for $i$-th parameter is computed by finding the $i$-th diagonal element of inverted FIM calculated using the expression \eqref{pro:FIMdiagProperty}. Note that all the decisions for frequency adjustment along the iterations are taken using the information about the FIM that is calculated using the estimated parameters' values.  As the set of frequencies is being modified for most of the parameters, the CRLB is improving. For parameters $\theta_7, \theta_8$ and $\theta_9$, the CRLB at the final set of frequencies is slightly, but negligibly increased, by $0.70\%$, $1.71\%$ and $0.05\%$, respectively. From the figure, we also conclude that by using the estimated parameters from each iteration to calculate the CRLB, the approximation is tracking well the true values.

\begin{figure}[H]
    \centering
    \includegraphics[scale = 0.5]{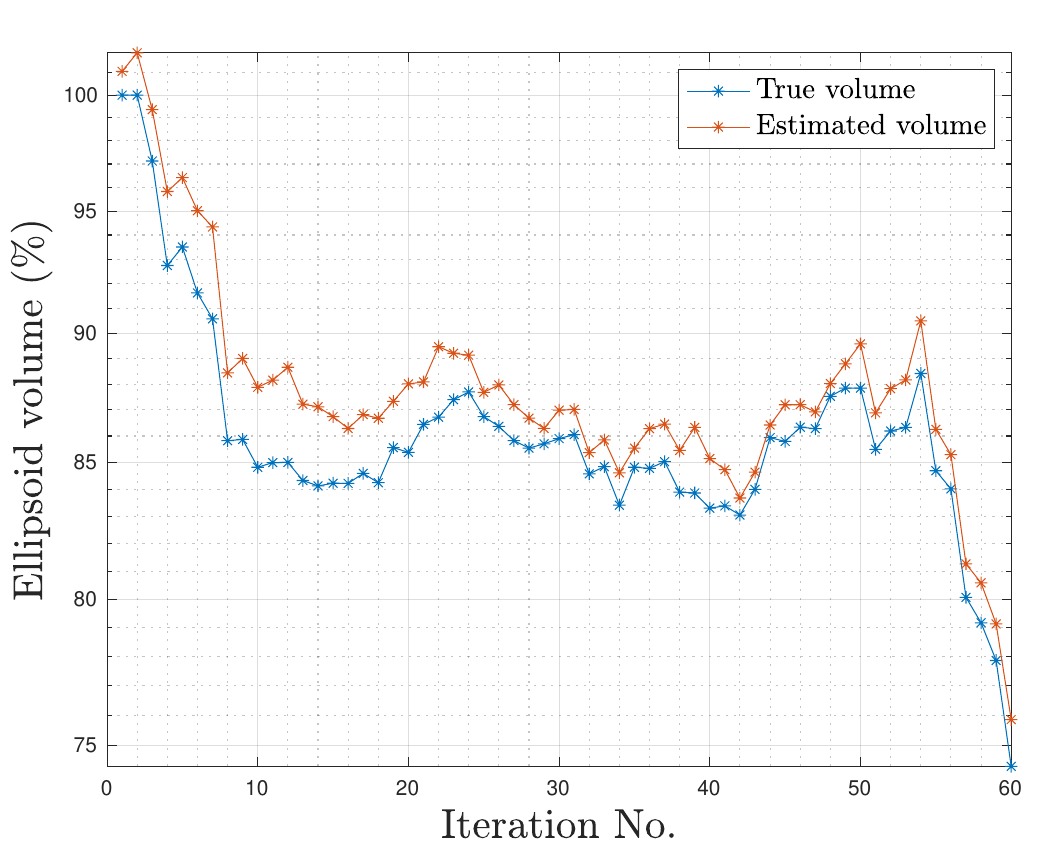}
    \caption{Evolution of the confidence ellipsoid volume while adjusting the frequencies along the algorithm's iterations.}
    \label{fig:EllipsoidVol}
\end{figure}

To quantify the overall improvement, in Fig.~\ref{fig:EllipsoidVol}, we also show the evolution of the parameters confidence ellipsoid's volume (normalized with the true initial volume of the confidence ellipsoid) along the iterations, calculated using \eqref{eq:ellipsoid_volume} and evaluated at both true and estimated parameters' values at each iteration. The frequency adjustment, while maximizing the lowest eigenvalue of the FIM and consequently minimizing the highest eigenvalue of its inverse, also reduces the volume of the confidence ellipse and improves the overall parameter estimation. After the termination of the algorithm, the confidence ellipsoid decreased in volume by at least $25\%$.

\pagebreak

\section{Conclusion}

In this paper, we presented a rigorous formulation of the ECM parameters identification problem for Li-ion batteries from EIS measurements. We developed the method to initialize the parameters of one of the most complete ECMs containing ten parameters. This automated method can be implemented into EIS measuring device software to help users run the CNLS method with a good starting point. We then studied and derived general CRLB applied to this specific identification problem, enabling us to design the experiments by quantifying the best possible accuracy of the parameter estimates. Since these values are highly dependent on the frequency set at which the EIS measurements are done, we developed the algorithm that modifies the initial logarithmic frequency span used for EIS measurements. This is done by utilizing the FIM and maximizing its lowest eigenvalue via the E-optimal design. As a result, the variances for most of the ECM parameters were significantly lowered. In the detailed numerical study,  we showed that, after executing our algorithm, parameters improved their variance by $14.34\%$ on average. The overall accuracy, quantified by the volume of the confidence ellipsoid, improved by at least $25\%$. The developed method for frequency adjustments for performing the EIS can also be incorporated within a measurement setup when characterizing the Li-ion cells. A similar approach can be potentially applied to other electrochemical systems which require impedance characterization and precise ECM identification.

\section*{Acknowledgements}
This research is carried out in the frame of EPFL – PSA Stationary Storage Study, financed by Stellantis and of the Swiss Circular Economy Model for Automotive Lithium Batteries (CircuBAT) Flagship project, with the financial support of the Swiss Innovation Agency (Innosuisse - Flagship Initiative) (FLAGSHIP PFFS-21-20).
\pagebreak

\clearpage
\bibliographystyle{ieeetr}
\bibliography{CRLB_Bibliography}

\begin{thebibliography}{10}

\bibitem{zhao_observability_2017}
S.~Zhao, S.~R. Duncan, and D.~A. Howey, ``Observability {Analysis} and {State} {Estimation} of {Lithium}-{Ion} {Batteries} in the {Presence} of {Sensor} {Biases},'' {\em IEEE Transactions on Control Systems Technology}, vol.~25, pp.~326--333, Jan. 2017.
\newblock Conference Name: IEEE Transactions on Control Systems Technology.

\bibitem{koseoglou_novel_2021}
M.~Koseoglou, E.~Tsioumas, D.~Papagiannis, N.~Jabbour, and C.~Mademlis, ``A {Novel} {On}-{Board} {Electrochemical} {Impedance} {Spectroscopy} {System} for {Real}-{Time} {Battery} {Impedance} {Estimation},'' {\em IEEE Transactions on Power Electronics}, vol.~36, pp.~10776--10787, Sept. 2021.
\newblock Conference Name: IEEE Transactions on Power Electronics.

\bibitem{alavi_identifiability_2017}
S.~M.~M. Alavi, A.~Mahdi, S.~J. Payne, and D.~A. Howey, ``Identifiability of {Generalized} {Randles} {Circuit} {Models},'' {\em IEEE Transactions on Control Systems Technology}, vol.~25, pp.~2112--2120, Nov. 2017.
\newblock Conference Name: IEEE Transactions on Control Systems Technology.

\bibitem{alavi_time-domain_2015}
S.~M.~M. Alavi, C.~R. Birkl, and D.~A. Howey, ``Time-domain fitting of battery electrochemical impedance models,'' {\em Journal of Power Sources}, vol.~288, pp.~345--352, Aug. 2015.

\bibitem{iurilli_use_2021}
P.~Iurilli, C.~Brivio, and V.~Wood, ``On the use of electrochemical impedance spectroscopy to characterize and model the aging phenomena of lithium-ion batteries: a critical review,'' {\em Journal of Power Sources}, vol.~505, p.~229860, Sept. 2021.

\bibitem{macdonald_analysis_1977}
J.~R. Macdonald and J.~A. Garber, ``Analysis of {Impedance} and {Admittance} {Data} for {Solids} and {Liquids},'' {\em Journal of The Electrochemical Society}, vol.~124, p.~1022, July 1977.
\newblock Publisher: IOP Publishing.

\bibitem{orazem_electrochemical_2017}
M.~E. Orazem and B.~Tribollet, {\em Electrochemical impedance spectroscopy}.
\newblock Hoboken, New Jersey: John Wiley \& Sons, Inc., 2017.
\newblock OCLC: 1104176375.

\bibitem{nocedal_numerical_2006}
J.~Nocedal and S.~J. Wright, {\em Numerical optimization}.
\newblock Springer series in operations research, New York: Springer, 2nd ed~ed., 2006.
\newblock OCLC: ocm68629100.

\bibitem{geuten_experimental_2007}
K.~Geuten, T.~Massingham, P.~Darius, E.~Smets, and N.~Goldman, ``Experimental {Design} {Criteria} in {Phylogenetics}: {Where} to {Add} {Taxa},'' {\em Systematic biology}, vol.~56, pp.~609--22, Sept. 2007.

\bibitem{goos_optimal_2011}
P.~Goos and B.~Jones, {\em Optimal design of experiments: a case study approach}.
\newblock Hoboken, N.J: Wiley, 2011.

\bibitem{pozzi_optimal_2018}
A.~Pozzi, G.~Ciaramella, K.~Gopalakrishnan, S.~Volkwein, and D.~M. Raimondo, ``Optimal {Design} of {Experiment} for {Parameter} {Estimation} of a {Single} {Particle} {Model} for {Lithiumion} {Batteries},'' in {\em 2018 {IEEE} {Conference} on {Decision} and {Control} ({CDC})}, (Miami Beach, FL), pp.~6482--6487, IEEE, Dec. 2018.

\bibitem{schmidt_experiment-driven_2010}
A.~P. Schmidt, M.~Bitzer, A.~W. Imre, and L.~Guzzella, ``Experiment-driven electrochemical modeling and systematic parameterization for a lithium-ion battery cell,'' {\em Journal of Power Sources}, vol.~195, pp.~5071--5080, Aug. 2010.

\bibitem{pillai_optimizing_2022}
P.~Pillai, S.~Sundaresan, K.~R. Pattipati, and B.~Balasingam, ``Optimizing {Current} {Profiles} for {Efficient} {Online} {Estimation} of {Battery} {Equivalent} {Circuit} {Model} {Parameters} {Based} on {Cramer}–{Rao} {Lower} {Bound},'' {\em Energies}, vol.~15, p.~8441, Nov. 2022.

\bibitem{rothenberger_genetic_2015}
M.~J. Rothenberger, D.~J. Docimo, M.~Ghanaatpishe, and H.~K. Fathy, ``Genetic optimization and experimental validation of a test cycle that maximizes parameter identifiability for a {Li}-ion equivalent-circuit battery model,'' {\em Journal of Energy Storage}, vol.~4, pp.~156--166, Dec. 2015.

\bibitem{du_information_2022}
X.~Du, J.~Meng, Y.~Zhang, X.~Huang, S.~Wang, P.~Liu, and T.~Liu, ``An {Information} {Appraisal} {Procedure}: {Endows} {Reliable} {Online} {Parameter} {Identification} to {Lithium}-{Ion} {Battery} {Model},'' {\em IEEE Transactions on Industrial Electronics}, vol.~69, pp.~5889--5899, June 2022.
\newblock Conference Name: IEEE Transactions on Industrial Electronics.

\bibitem{macdonald_flexible_1987}
J.~R. Macdonald and L.~D. Potter, ``A flexible procedure for analyzing impedance spectroscopy results: {Description} and illustrations,'' {\em Solid State Ionics}, vol.~24, pp.~61--79, June 1987.

\bibitem{abaspour_robust_2022}
M.~Abaspour, K.~R. Pattipati, B.~Shahrrava, and B.~Balasingam, ``Robust {Approach} to {Battery} {Equivalent}-{Circuit}-{Model} {Parameter} {Extraction} {Using} {Electrochemical} {Impedance} {Spectroscopy},'' {\em Energies}, vol.~15, p.~9251, Dec. 2022.

\bibitem{ospina_agudelo_identification_2019}
B.~Ospina~Agudelo, W.~Zamboni, E.~Monmasson, and G.~Spagnuolo, ``Identification of battery circuit model from {EIS} data,'' (Saint Pierre d’Oléron, France), hal-02915697, June 2019.

\bibitem{troltzsch_characterizing_2006}
U.~Tröltzsch, O.~Kanoun, and H.-R. Tränkler, ``Characterizing aging effects of lithium ion batteries by impedance spectroscopy,'' {\em Electrochimica Acta}, vol.~51, pp.~1664--1672, Jan. 2006.

\bibitem{islam_unification_2020}
S.~M.~R. Islam, S.-Y. Park, and B.~Balasingam, ``Unification of {Internal} {Resistance} {Estimation} {Methods} for {Li}-{Ion} {Batteries} {Using} {Hysteresis}-{Free} {Equivalent} {Circuit} {Models},'' {\em Batteries}, vol.~6, p.~32, June 2020.

\bibitem{wu_battery_2023}
Y.~Wu, S.~Sundaresan, and B.~Balasingam, ``Battery {Parameter} {Analysis} through {Electrochemical} {Impedance} {Spectroscopy} at {Different} {State} of {Charge} {Levels},'' {\em Journal of Low Power Electronics and Applications}, vol.~13, p.~29, Apr. 2023.

\bibitem{wang_electrochemical_2021}
S.~Wang, J.~Zhang, O.~Gharbi, V.~Vivier, M.~Gao, and M.~E. Orazem, ``Electrochemical impedance spectroscopy,'' {\em Nature Reviews Methods Primers}, vol.~1, p.~41, June 2021.

\bibitem{lerro_tracking_1993}
D.~Lerro and Y.~Bar-Shalom, ``Tracking with debiased consistent converted measurements versus {EKF},'' {\em IEEE Transactions on Aerospace and Electronic Systems}, vol.~29, pp.~1015--1022, July 1993.
\newblock Conference Name: IEEE Transactions on Aerospace and Electronic Systems.

\bibitem{milano_static_2016}
M.~Paolone, J.-Y. Le~Boudec, S.~Sarri, and L.~Zanni, ``Static and recursive {PMU}-based state estimation processes for transmission and distribution power grids,'' in {\em Advances in {Power} {System} {Modelling}, {Control} and {Stability} {Analysis}} ({Milano}, ed.), pp.~189--239, Institution of Engineering and Technology, Sept. 2016.

\bibitem{kay_fundamentals_1993}
S.~M. Kay, {\em Fundamentals of statistical signal processing}.
\newblock Prentice {Hall} signal processing series, Englewood Cliffs, N.J: Prentice-Hall PTR, 1993.

\bibitem{nielsen_elementary_2018}
F.~Nielsen, ``An {Elementary} {Introduction} to {Information} {Geometry},'' {\em Entropy (Basel, Switzerland)}, vol.~22, Aug. 2018.

\bibitem{wilson_volume_2009}
A.~J. Wilson, ``Volume of n-dimensional ellipsoid,'' {\em Sciencia Acta Xaveriana}, vol.~1, no.~1, pp.~101--106, 2009.

\end{thebibliography}

\end{document}